\documentclass[12pt]{article}
\usepackage{amssymb,amsmath,amscd}
\makeatletter

\@addtoreset{equation}{section}
\def\section{\@startsection {section}{1}{\z@}{-2.25ex plus -1ex minus
 -.2ex}{1.0ex plus .2ex}{\large\bf}}
\def\subsection{\@startsection{subsection}{2}{\z@}{-2.0ex plus%
 -1ex minus -.2ex}{0.5ex plus .2ex}{\bf}}

\def\Ad{\mathrm{Ad}}
\def\ad{\mathrm{ad}}
\newcommand{\inv}[0]{{-1}}

\newcommand{\cif}[0]{\mathcal{C}^\infty}

\def\bfh{{\mbox{\boldmath $f$}}}

\def\ba{{\mbox{\boldmath $a$}}}
\def\br{{\mbox{\boldmath $r$}}}
\def\bx{{\mbox{\boldmath $x$}}}

\def\by{{\mbox{\boldmath $y$}}}
\def\bj{{\mbox{\boldmath $j$}}}
\def\bq{{\mbox{\boldmath $q$}}}
\def\bc{{\mbox{\boldmath $c$}}}

\def\bk{{\mbox{\boldmath $k$}}}
\def\bp{{\mbox{\boldmath $p$}}}

\def\bq{{\mbox{\boldmath $q$}}}

\def\bn{{\mbox{\boldmath $n$}}}
\def\bl{{\mbox{\boldmath $l$}}}

\newcommand{\ZZ}{\mathbb{Z}}

\newcommand{\RR}{\mathbb{R}}
\newcommand{\CC}{\mathbb{C}}

\def\Poi{ P_3^\uparrow }
\def\Lor{L_3^\uparrow}

\newcommand{\hi}[0]{{H_i}}

\newcommand{\ai}[0]{{A_i}}
\newcommand{\bi}[0]{{B_i}}

\newcommand{\hyp}[0]{\mathbb{H}^2}

\newtheorem{theorem}{Theorem}[section]

\newtheorem{corollary}[theorem]{Corollary}

\def\bea{\begin{eqnarray}}
\def\eea{\end{eqnarray}}
\def\bmz{\left(\begin{array}{2,2}}
\def\emz{\end{array}\right)}
\def\bmd{\left(\begin{array}{3,3}}
\def\emd{\end{array}\right)}

\begin{document}
\parskip 5pt
\parindent 0pt

\begin{center}
\baselineskip 24 pt {\Large \bf Grafting and Poisson structure in
(2+1)-gravity with vanishing cosmological constant}

\baselineskip 18 pt

\vspace{1cm} {C.~Meusburger}\footnote{\tt  cmeusburger@perimeterinstitute.ca}\\
Perimeter Institute for Theoretical Physics\\
31 Caroline Street North,
Waterloo, Ontario N2L 2Y5, Canada\\

\vspace{0.5cm}

{31 July 2005}

\end{center}

\begin{abstract}
\noindent We relate the geometrical construction of
(2+1)-spacetimes via grafting to phase space and Poisson structure
in the Chern-Simons formulation of (2+1)-dimensional gravity with
vanishing cosmological constant on manifolds of topology
$\RR\times S_g$, where $S_g$ is an orientable two-surface of genus
$g>1$. We show how grafting along simple closed geodesics
$\lambda$ is implemented in the Chern-Simons formalism and derive
explicit expressions for its action on the holonomies of general
closed curves on $S_g$.  We prove that this action is generated
via the Poisson bracket by a gauge invariant observable associated
to the holonomy of $\lambda$. We deduce a symmetry
relation between the Poisson brackets of observables associated to
the Lorentz and translational components of the holonomies of
general closed curves on $S_g$ and discuss its physical
interpretation. Finally, we relate the action of grafting on the
phase space to the action of Dehn twists and show that grafting
can be viewed as a Dehn twist with a formal parameter $\theta$
satisfying $\theta^2=0$.

\baselineskip 12pt \noindent
\end{abstract}

\section{Introduction}

(2+1)-dimensional gravity is of physical interest as a toy model
for the (3+1)-dimensional case. It is used as a testing ground
which allows one to investigate conceptual questions arising in
the quantisation of gravity without being hindered by the
technical complexity in higher dimensions. One of  these questions
is the problem of "quantising geometry" or, more concretely, the
problem of recovering geometrical objects with a clear physical
interpretation from the gauge theory-like formulations  used as a
starting point for quantisation.

In (2+1)-dimensions, the relation between Einstein's theory of
gravity and gauge theory is more direct than in higher dimensional
cases, since the theory  takes the form of a Chern-Simons gauge
theory. Depending on the value of the cosmological constant,
vacuum solutions of Einstein's equations of motion are flat or of
constant curvature. The theory has only a finite number of
physical degrees of freedom arising from the matter content and
topology of the spacetime. This absence of local gravitational
degrees of freedom manifests itself mathematically in the
possibility to formulate the theory as a Chern-Simons gauge theory
\cite{AT,Witten1} where the gauge group is the (2+1)-dimensional
Poincar\'e group $\Poi$, the group $SO(3,1)\cong
SL(2,\mathbb{C})/\mathbb{Z}_2$ or $SO(2,2)\cong (SL(2,\RR)\times
SL(2,\RR))/\mathbb{Z}_2$, respectively, for cosmological constant
$\Lambda=0$, $\Lambda>0$ and $\Lambda<0$.

The main advantage of the Chern-Simons formulation of
(2+1)-dimensional gravity is that it allows one to apply gauge
theoretical concepts and methods which give rise to an efficient
description of phase space and Poisson structure. As Einstein's
equations of motion take the form of a flatness condition on the
gauge field, physical states can be characterised by holonomies,
and conjugation invariant functions of the holonomies form a
complete set of physical observables. Starting with the work of
Nelson and Regge \cite{RN,RN1,RN2,RN3,RN4}, Martin \cite{martin}
and of Ashtekar, Husain, Rovelli, Samuel and Smolin \cite{ahrss},
the description of (2+1)-dimensional gravity in terms of
holonomies and the associated gauge invariant observables has
proven useful in clarifying the structure of its classical phase
space as well as in quantisation. An overview of different
approaches and results is given in \cite{Carlipbook}.

The disadvantage of this approach is that it makes it difficult to
recover the geometrical  picture of a spacetime manifold and
thereby complicates the physical interpretation of the theory.
Except for cases where the holonomies take a particularly simple
form such as static spacetimes and the torus universe, it is in
general not obvious how the description of the phase space in
terms of holonomies and associated gauge invariant observables
gives rise to a Lorentz metric on a spacetime manifold. The first
to address this problem for general spacetimes
  was Mess \cite{mess}, who showed how the geometry of
(2+1)-dimensional spacetimes  can be reconstructed from a set of
holonomies.  More recent results on this problem are obtained in
the papers by Benedetti and Guadagnini \cite{bg} and by Benedetti
and Bonsante \cite{bb}, which are going to be our main references.
They describe the construction of evolving spacetimes from static
ones via the geometrical procedure of grafting, which,
essentially, consists of inserting small annuli along certain
geodesics of the spacetime. As they establish a unified picture
for all values of the cosmological constant and show how this
change of geometry affects the holonomies, they  clarify the
relation between holonomies and spacetime geometry considerably.

However, despite these results, the problem of relating spacetime
geometry and the description of phase space and Poisson structure
in terms of holonomies has not yet been fully solved. The missing
link is the role of the Poisson structure. A complete
understanding of the gauge invariant observables must include a
physical interpretation of the transformations on phase space they
generate via the Poisson bracket. Conversely, to interpret the
geometrical construction of evolving (2+1)-spacetimes via grafting
as a physical transformation, one needs to determine how it
affects phase space and Poisson structure.

This  paper addresses these questions  for (2+1)-gravity with
vanishing cosmological constant on manifolds of topology
$\RR\times S_g$, where $S_g$ is an orientable two-surface of genus
$g>1$. It relates the construction of evolving (2+1)-spacetimes
via grafting along simple, closed curves to the description of the
phase space in terms of holonomies and the associated gauge
invariant observables. The main results can be stated as follows.
\begin{enumerate}

\item We show how grafting along a closed, simple geodesic
 is implemented in the
Chern-Simons formulation of (2+1)-dimensional gravity. Using the
parametrisation of the phase space in terms of holonomies given in
\cite{we1,we2}, we deduce explicit expressions for the action of
grafting on the holonomies of general curves on $S_g$ and
investigate its properties as a transformation on phase space.

\item We derive the Hamiltonian that generates
 this grafting transformation via the Poisson bracket. This
 Hamiltonian is one of the two basic gauge invariant observables
 associated to a closed curve on $S_g$ and obtained from the Lorentz component of its
 holonomy.

\item We demonstrate  that there is a  symmetry relation between
the transformation of the observables associated to a curve $\eta$
under grafting along $\lambda$ and the transformation of the
corresponding observables for $\lambda$ under
 grafting along $\eta$. Infinitesimally, this relation takes the
 form of a general identity for the Poisson
 brackets of certain observables associated to the two curves.

\item We  show that the action of grafting in our description of
the phase space is closely related to the action of
(infinitesimal) Dehn twists investigated in an earlier paper
\cite{we3}. Essentially, grafting can be viewed as a Dehn twist
with a formal parameter $\theta$ satisfying $\theta^2=0$.
\end{enumerate}

The paper is structured as follows. In Sect.~\ref{bgsect} we
introduce the relevant definitions and notation, present some
background on the (2+1)-dimensional Poincar\'e group and on
hyperbolic geometry and summarise the description of grafted
(2+1)-spacetimes  in \cite{bg,bb} for the case of grafting along
multicurves.

In Sect.~\ref{CSbg}, we briefly review the Hamiltonian version of
the Chern-Simons formulation of (2+1)-dimensional gravity. We
discuss the role of holonomies and summarise the relevant results
of \cite{we1,we2}, in which phase space and Poisson structure are
 characterised by a symplectic potential on the manifold
$(\Poi)^{2g}$ with different copies of $\Poi$ standing for the
holonomies of a set of generators of the fundamental group
$\pi_1(S_g)$.

Sect.~\ref{Csgraft} discusses the implementation of grafting along
closed, simple geodesics in the Chern-Simons formalism. We show
how the geometrical procedure of grafting in \cite{bg,bb} gives
rise to a transformation on the extended phase space $(\Poi)^{2g}$
and derive formulas for its action on the holonomies of general
elements of the fundamental group $\pi_1(S_g)$.

Sect.~\ref{massgen} establishes the relation of grafting and
Poisson structure. After expressing the symplectic potential on
the extended phase space $(\Poi)^{2g}$ in terms of variables
adapted to the grafting transformations, we show that these
transformations are generated by gauge invariant Hamiltonians and
therefore act as Poisson isomorphisms. We deduce a general
symmetry relation between the Poisson
 brackets of certain observables associated to general closed
 curves on $S_g$.

In Sect.~\ref{grdtrel}, we explore the link between grafting and
Dehn twists. We review the results concerning Dehn twists
 derived in \cite{we3} and introduce a graphical procedure which
 allows one to determine the action of grafting on the
 holonomies of general closed curves on $S_g$. By means of this procedure, we then
 demonstrate that there is a close relation between the
 action of grafting and (infinitesimal) Dehn twists.

In  Sect.~\ref{exsect} we illustrate the general results from
Sect.~\ref{Csgraft} to \ref{grdtrel}   by applying them to a
concrete example. Sect.~\ref{conc} contains a summary of our
results and concluding remarks.

\section{Grafted (2+1) spacetimes with vanishing cosmological constant: the geometrical viewpoint}
\label{bgsect}

\subsection{The (2+1)-dimensional Poincar\'e group}
\label{notdef}

Throughout the paper we use Einstein's summation convention.
Indices are raised and lowered with the three-dimensional
Minkowski metric $\eta=\text{diag}(1,-1,-1)$, and $\bx\cdot\by$
stands for $\eta(\bx,\by)$.

In the following $\Lor$ and $\Poi=\Lor\ltimes\RR^3$ denote,
respectively, the the (2+1)-dimensional proper orthochronous
Lorentz and Poincar\'e group. We identify $\RR^3$ and the Lie
algebra $so(2,1)=\text{Lie}\;\Lor$ as vector spaces.  The action
of $\Lor$ on $\RR^3$ in its matrix representation then agrees with
its action on $so(2,1)$ via the adjoint action
\begin{align}
\label{lieident} \Ad(u)\bp=p^a\,u J_a u^\inv=u^{b}_{\;\;a}p^a
J_b\qquad \bp=(p^0,p^1,p^2)\cong p^a J_a,
\end{align}
where $J_a$, $a=0,1,2$, are the generators of $so(2,1)$. For
notational consistency with earlier papers \cite{we1,we2,we3}
considering
 more general gauge groups  we will use the
notation $\Ad(u)\bp$ throughout the paper and often do not
distinguish notationally between elements of $so(2,1)$ and
associated vectors in $\RR^3$. With the parametrisation \bea
\label{gparam} (u,\ba)=(u,-\Ad(u)\bj)\in\Poi\qquad\qquad
u\in\Lor,\; \ba,\bj\in\mathbb{R}^3, \eea the group multiplication
in $\Poi$ is then given by \bea \label{groupmult}
 (u_1,\ba_1)\cdot(u_2,\ba_2)=(u_1\cdot
 u_2,\ba_1+\Ad(u_1)\ba_2)=(u_1\cdot u_2,-\Ad(u_1u_2)(\bj_2+\Ad(u_2^\inv)\bj_1) ).\eea
The Lie algebra of $\Poi$ is $\text{Lie}\; \Poi=iso(2,1)$.
Denoting by  $J_a$, $a=0,1,2$, the generators of $so(2,1)$ by
 $P_a$, $a=0,1,2$, the generators of the translations, and
 choosing the  convention $\epsilon_{012}=1$ for the epsilon
 tensor,
we have the Lie bracket
 \bea
\label{poinccomm}
 [P_a,P_b] =0,
\quad[J_a,J_b]=\epsilon_{abc} J^c,\quad
[J_a,P_b]=\epsilon_{abc}P^c, \eea
 and a non-degenerate,
$\Ad$-invariant bilinear form $\langle\,,\,\rangle$ on $iso(2,1)$
is given by
\begin{align} \label{inprod} \langle J_a, P^b\rangle = \delta_a^b,
\quad \langle J_a, J_b\rangle = \langle P^a,P^b\rangle = 0.
\end{align}

We represent the generators of $so(2,1)$ by the matrices
\begin{align}
\label{jdef} (J_a)_{bc}=-\epsilon_{abc}
\end{align}
and denote by  $\exp: \;so(2,1)\rightarrow \Lor,\; p^a J_a\mapsto
e^{p^a J_a}$ the exponential map for $\Lor$.  As this map is
surjective, see for example \cite{grig,munds}, elements of $\Lor$
can be parametrised in terms of a vector $\bp\in\RR^3$ with
$p^0\geq 0$ as
$$
\label{lorpar} u=e^{-p^a J_a}.
$$
Using expression \eqref{jdef} for the generators of $so(2,1)$ and
setting \begin{align} \label{phatdef}
\hat\bp=\tfrac{1}{m}\bp\qquad\text{for}\; m^2:=|\bp^2|\neq 0,
\end{align} we find
\begin{align} \label{adjform} & u_{ab}=\begin{cases}\hat p_a
\hat p_b+\cos m(\eta_{ab}-\hat p_a\hat
p_b)+\sin m \epsilon_{abc}\hat p^c & p^ap_a=m^2>0\\
\eta_{ab}+\epsilon_{abc}p^c+\tfrac{1}{2}p_a p_b & p^a p_a=0\\
-\hat p_a \hat p_b+\cosh m(\eta_{ab}+\hat p_a\hat p_b)+\sinh m
\epsilon_{abc}\hat p^c & p^ap_a=-m^2<0.
\end{cases}
\end{align}
Elements $u=e^{-p^a J_a}\in\Lor$ are called elliptic, parabolic
and hyperbolic, respectively,  if $\bp^2>0$, $\bp^2=0$ and
$\bp^2<0$. Note that the exponential map is not injective, since
 $e^{2\pi n J_0}=1$
for $n\in\ZZ$. However, in this paper  we will be concerned with
hyperbolic elements, for which the parametrisation \eqref{adjform}
in terms of a spacelike vector $\bp=(p^0,p^1,p^2)$ with $p^0\geq
0$ is unique.

A convenient way of describing the Lie algebra $iso(2,1)$ and the
group $\Poi$ has been introduced in \cite{martin}. It relies on a
formal parameter $\theta$ with $\theta^2=0$ analogous to the one
occurring in supersymmetry. With the definition
\begin{align}
\label{martinp} (P_a)_{bc}=\theta (J_a)_{bc}=-\theta
\epsilon_{abc},
\end{align}
it follows that the commutator of the matrices  $P_a,J_a$,
$a=0,1,2$, is the Lie bracket \eqref{poinccomm} of the
(2+1)-dimensional Poincar\'e algebra.  Definition \eqref{martinp}
also allows one to parametrise elements of the group $\Poi$.
Identifying
\begin{align}
\label{mident} (u,\ba)\cong(1+\theta a^bJ_b)u,
\end{align}
one obtains the multiplication law
\begin{align}
(1+ \theta a_1^b J_b)u_1\cdot (1+\theta a_2^cJ_c)u_2&=u_1u_2
+\theta a_1^b J_b u_1u_2+\theta u_1 a_2^b J_b u_2 +\theta^2 a_1^b
J_b u_1
a_2^c J_c u_2\\
&=(1+\theta(a_1^b+u_1a_2^bJ_b u_1^\inv))u_1u_2,\nonumber
\end{align}
and with the identification \eqref{lieident} of $so(2,1)$ and
$\RR^3$ one recovers the group multiplication law
\eqref{groupmult}. Furthermore, the introduction of the parameter
$\theta$ makes it possible to express the exponential map $\exp:\;
iso(2,1)\rightarrow\Poi$ in terms of the exponential map $\exp:
so(2,1)\rightarrow\Lor$ for the (2+1)-dimensional Lorentz group by
setting
\begin{align}
\label{martinexp} e^{p^a J_a+k^a P_a}=e^{(p^a+\theta
k^a)J_a}=\sum_{n=0}^\infty \frac{(p^a
J_a)^n}{n!}+\theta\sum_{n=0}^\infty\sum_{m=0}^{n-1} \frac{(p^a
J_a)^m (k^b J_b)(p^c
J_c)^{n-m-1}\!\!\!\!\!\!\!\!\!\!\!\!\!\!\!}{n!} .
\end{align}
To link the parametrisation of elements of $\Poi$ in terms of the
exponential map with the parametrisation \eqref{gparam}
\begin{align}
\label{kexp} (u,-\Ad(u)\bj)=e^{-(p^a+\theta k^a) J_a}\qquad
u\in\Lor,\bj\in\RR^3, (p^a+\theta k^a)J_a\in iso(2,1),
\end{align}
one uses the identity
\begin{align}
\label{tident} [(p^aJ_a)^n, k^b J_b]=\sum_{m=1}^n
\left(\begin{array}{l} n\\ m\end{array}\right) \ad_{p^aJ_a}^m (k^b
J_b)\cdot (p^c J_c)^{n-m}
\end{align}
in \eqref{martinexp} and finds that the elements $u\in\Lor$,
$\bj\in\RR^3$ are given by
\begin{align}
\label{martinhelp} &u=e^{-p^a J_a},
\bj=T(\bp)\bk\quad\text{with}\;T:\RR^3\rightarrow \RR^3\\
\label{tdef} &T(\bp)^{ab}k_b J_a=\sum_{n=0}^\infty \frac{\ad_{p^a
J_a}^n(k^a J_a)}{(n+1)!}=k^a J_a+\tfrac{1}{2}[p^b J_b, k^a
J_a]+\tfrac{1}{6}[p^c J_c,[p^b J_b, k^a J_a]]+\ldots.
\end{align}
Note that the linear map $T(\bp)$ is the same as the one
considered in \cite{we1, we2}, where its properties are discussed
in more detail. In particular, it is shown that $T(\bp)$ is
bijective, maps $\bp$ to itself and satisfies $\Ad(e^{-p^a
J_a})T(\bp)=T(-\bp)$. Its inverse $T^{-1}(\bp):\RR^3\rightarrow
\RR^3$ plays an important role in the parametrisation of the
right- and left-invariant vector fields $J^L_a, J^R_a$ on $\Lor$.
For any $F\in \cif(\Lor)$, we have
\begin{align}
\label{lorvecfs} &J_a^L F(e^{-p^b
J_b})=\frac{d}{dt}|_{t=0}F(e^{-tJ_a} e^{-p^b
J_b})=T^\inv(\bp)^{ab}\frac{\partial F}{\partial p^b}\\
&J_a^R F(e^{-p^b J_b})=\frac{d}{dt}|_{t=0}F(e^{-p^b
J_b}e^{tJ_a})=-\Ad(e^{p^b
J_b})^{a}_{\;\;c}T^\inv(\bp)^{cb}\frac{\partial F}{\partial
p^b}=-T^\inv(-\bp)^{ab}\frac{\partial F}{\partial p^b}.\nonumber
\end{align}

\subsection{Hyperbolic geometry}
\label{hypgeom}

In this subsection we summarise some facts from hyperbolic
geometry, mostly following the presentation in \cite{bp}. For a
more specialised treatment focusing on Fuchsian groups see also
\cite{fgroups}.

In the following we denote by $H_T\subset \RR^3$ the hyperboloids
of curvature $-\frac{1}{T}$ with the metric induced by the
(2+1)-dimensional Minkowski metric
\begin{align}
\label{hyper} H_T=\{\bx\in\RR^3\;|\, \bx^2=T^2,\;x^0>0\}
\end{align}
and realise hyperbolic space $\hyp$ as the hyperboloid
$\mathbb{H}^2=H_1$. The tangent plane in a point $\bp\in H_T$ is
given by
\begin{align}
T_{\!_{\footnotesize \bp}} H_T=\bp^\bot=\{\bx\in
\mathbb{R}^3\;|\;\bx\cdot\bp=0\},
\end{align}
and geodesics on $H_T$ are of the form
\begin{align}
\label{geods} \bc_{p,q}(t)=\bp\cosh t+\bq\sinh
t\qquad\text{with}\; \bp^2=T^2,\,\bq^2=-T^2,\,\bp\cdot \bq=0.
\end{align}
They are given as the intersection of $H_T$ with planes through
the origin, which can be characterised in terms of their unit
(Minkowski) normal vectors
\begin{align}
\label{normvect} \bc_{p,q}=H_T\cap
\bn_{p,q}^\bot\qquad\text{with}\qquad\bn_{p,q}=\tfrac{1}{T^2}\bp\times\bq
\in T_{\!_{\bc_{p,q}(t)}}H_T\;\forall t\in\RR.
\end{align}
The isometry group of the hyperboloids $H_T$ is the
(2+1)-dimensional proper orthochronous Lorentz group $\Lor$. The
subgroup stabilising a given geodesic maps the associated plane to
itself and is generated by the plane's normal vector. More
precisely, for a geodesic $\bc_{p,q}$ parametrised as in
\eqref{geods} and with associated normal vector $\bn_{p,q}$ as in
\eqref{normvect}, one has
\begin{align}
\label{adgeod} \Ad(e^{ \alpha n_{p,q}^a J_a})
\bc_{p,q}(t)=&\cosh(t+ \alpha)\bp+\sinh(t+
\alpha)\bq\qquad\alpha\in\RR.
\end{align}

The uniformization theorem implies that any compact, oriented
two-manifold of genus $g>1$ with a metric of constant negative
curvature  is given as a quotient $S_\Gamma=H_T/\Gamma$ of a
hyperboloid $H_T$ by the action of a cocompact Fuchsian group
 with $2g$ hyperbolic generators
\begin{align}
\label{Fuchsgroup} \Gamma=\langle v_{A_1},v_{B_1},\ldots,
v_{A_g},v_{B_g}; [v_{B_g},
v_{A_g}^\inv]\cdots[v_{B_1},v_{A_1}^\inv]=1\rangle\subset\Lor.
\end{align}
The group $\Gamma$ is isomorphic to the fundamental group
$\pi_1(S_\Gamma)$, and its action on the hyperboloid $H_T$ agrees
with the action of $\pi_1(S_\Gamma)$ via deck transformations. Via
its action on $H_T$, it induces a tesselation of $H_T$ by its
fundamental regions which are geodesic arc $4g$-gons. In
particular, there exists a geodesic arc $4g$-gon $P_\Gamma^T$ in
the tesselation of $H_T$, in the following referred to as
fundamental polygon, such that each of the generators of $\Gamma$
and their inverses map the polygon to one of its $4g$ neighbours.
If one labels the sides of the polygon by $a_1,
a_1',\ldots,b_1,b'_1,\ldots, a_g,a_g',b_g,b_g'$ as in
Fig.~\ref{poly1},  it follows that the generators of $\Gamma$ map
side $x\in\{a_1,\ldots, b_g\}$ of the polygon $P_\Gamma$ into
$x'\in\{a'_1,\ldots, b'_g\}$
\begin{align}
\label{fuchsels} \Ad(v_{A_i}): a_i\rightarrow a_i'\qquad
\Ad(v_{B_i}): b_i\rightarrow b_i'.
\end{align}
For a general polygon $P'$ in the tesselation related to
$P_\Gamma^T$ via $P'=\Ad(v) P_\Gamma^T$, $ v\in\Gamma$, the
elements of $\Gamma$ mapping this polygon into its $4g$ neighbours
are given by $ vv_{A_1}^{\pm 1} v^\inv,\ldots, v v_{B_g}^{\pm 1}
v^\inv$.

Geodesics on $S_\Gamma$ are the images of geodesics on $H_T$ under
 the projection $\Pi_\Gamma^T: H_T\rightarrow S_\Gamma$.
In particular, a geodesic $\bc_{p,q}$ on $H_T$ gives rise to a
closed geodesic on $S_\Gamma$ if and only if there exists a
nontrivial element $ \tilde v\in\Gamma$, the geodesic's holonomy,
which maps $\bc_{p,q}$ to itself. From \eqref{adgeod} it then
follows that the group element $\tilde v\in\Gamma$ is obtained by
exponentiating a multiple of the geodesic's normal vector
\begin{align}
\exists \alpha\in\RR^+:\qquad \tilde v=e^{\alpha n_{p,q}^a J_a}.
\end{align}
Closed geodesics on $S_\Gamma$ are therefore in one to one
correspondence with elements of the group $\Gamma$ and hence with
elements of the fundamental group $\pi_1(S_\Gamma)$. In the
following we will often not distinguish notationally between an
element of the fundamental group $\pi_1(S_\Gamma)$ and a closed
geodesic or a general closed curve on $S_\Gamma$ representing this
element.

\subsection{Grafting}
\label{graft}

Grafting along simple geodesics was first investigated in the
context of complex projective structures and Teichm\"uller theory
\cite{gol,hej,msk}. Following the work of Thurston \cite{th1, th2}
who considered general geodesic laminations, the topic has
attracted much interest in mathematics, for historical remarks see
for instance \cite{mcmull}. The role of geodesic laminations in
(2+1)-dimensional gravity was first explored by Mess \cite{mess}
who investigated the construction of (2+1)-dimensional spacetimes
from  a set of holonomies. More recent work on grafting in the
context of (2+1)-dimensional gravity  are the papers  by Benedetti
and Guadagnini \cite{bg} and by Benedetti and Bonsante \cite{bb}.
As we investigate a rather specific situation, namely grafting
along closed, simple geodesics in (2+1)-dimensional gravity with
vanishing cosmological constant $\Lambda$, we limit our
presentation to a summary of the grafting procedure described in
\cite{bg,bb} for the case of $\Lambda=0$ and multicurves. For a
more general treatment and a discussion of the relation between
this grafting procedure and grafting on the space of complex
projective structures, we refer the reader to \cite{bg,bb}.

Given a cocompact Fuchsian group $\Gamma$ with $2g$ generators,
there is
 a well-known procedure for constructing a flat (2+1)-dimensional
 spacetime of genus $g$ associated to this group, see  for example \cite{Carlipbook}.
One foliates the interior of the forward lightcone with the tip at
the origin by hyperboloids $H_T$. The cocompact Fuchsian group
$\Gamma$ acts on each hyperboloid $H_T$ and induces a tesselation
of $H_T$ by geodesic arc $4g$-gons which are mapped into each
other by the elements of $\Gamma$. The asssociated spacetime of
genus $g$ is then obtained by identifying on each hyperboloid the
points related by the action of $\Gamma$. It is shown in
\cite{Carlipbook} that the $\Poi$-valued holonomies of all curves
in the resulting spacetime have vanishing translational components
and lie in the subgroup $\Lor\subset \Poi$. However, these
spacetimes are of limited physical interest because they are
static \cite{Carlipbook}.

Grafting along measured geodesic laminations is a procedure which
allows one to construct non-static or genuinely evolving
(2+1)-spacetimes associated to a Fuchsian group $\Gamma$. In the
following we consider measured geodesic laminations which are
weighted multicurves, i.~e.~ countable or finite sets
\begin{align}
\label{multgamma} G_I^\Gamma=\{(c_i^\Gamma,w_i)\;|\;i\in I\}
\end{align}
of closed, simple non-intersecting geodesics $c_i^\Gamma$ on the
associated two-surface $S_\Gamma=\hyp/\Gamma$, each equipped with
a positive number, the weight $w_i>0$.
\begin{figure}
\vskip .3in \protect\input epsf \protect\epsfxsize=12truecm
\protect\centerline{\epsfbox{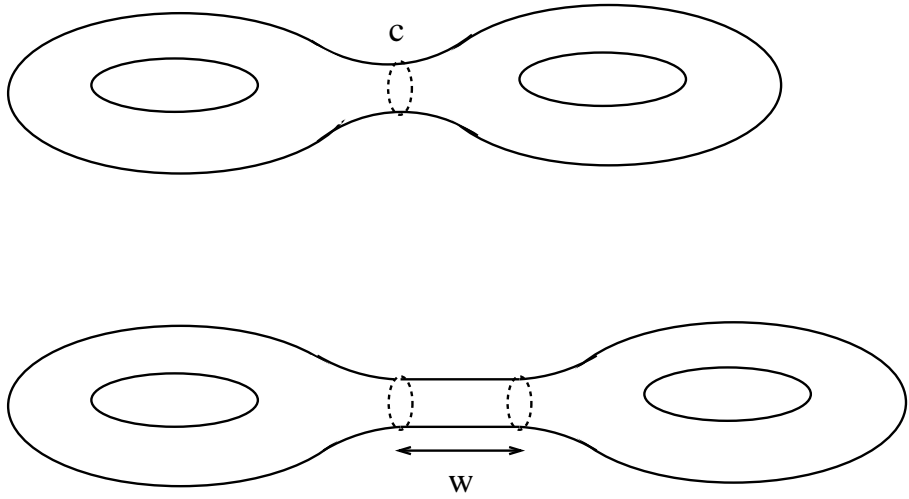}} \caption{Grafting along a
closed simple geodesic $c$ with weight $w$ on a genus 2 surface.}
\label{graft}
\end{figure}

Geometrically, grafting amounts to cutting the surface $S_\Gamma$
along each  geodesic $c_i$  and inserting a strip of width $w_i$
as indicated in Fig.~\ref{graft}.
\begin{figure}
\vskip .3in \protect\input epsf \protect\epsfxsize=12truecm
\protect\centerline{\epsfbox{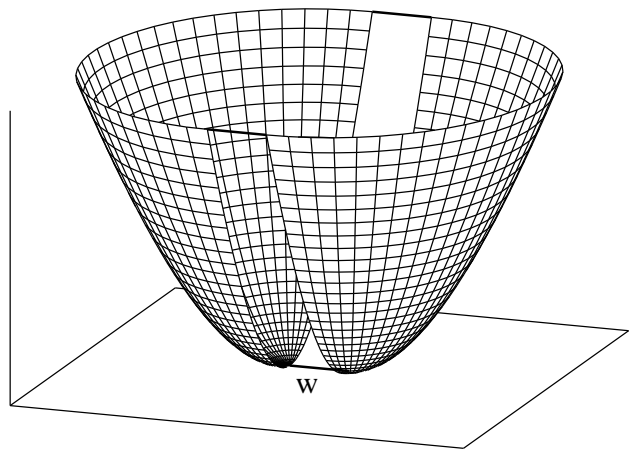}} \caption{Grafting along a
 geodesic with weight $w$ in hyperbolic space.} \label{hyper}
\end{figure}
 Equivalently, the construction can be described in
the universal cover $\hyp$. By lifting each geodesic $c_i^\Gamma$
to a geodesic $\bc_i$ on $\hyp$ and acting on it  with the
Fuchsian group $\Gamma$, one obtains a $\Gamma$-invariant
multicurve on $\hyp$
\begin{align}
\label{multlift} G_I=\{(\Ad(v)\bc_i,w_i)\,|\, i\in I,
v\in\Gamma\}.
\end{align}
One then cuts the hyperboloid  $\hyp$ along each geodesic $c_i$ in
$G_I$, shifts the resulting pieces in the direction of the
geodesics' normal vectors and glues in a strip of width $w_i$ as
shown in Fig.~\ref{hyper}. The cocompact Fuchsian group $\Gamma$
acts on the resulting surface in such a way that it identifies the
images of the points related by the canonical action of $\Gamma$
on $\hyp$, and the associated grafted genus $g$ surface is
obtained by taking the quotient with respect to this action of
$\Gamma$.

In the construction of flat (2+1)-spacetimes of topology
$\RR\times S_g$ via grafting, the grafting procedure is performed
for each value of the time  parametrising $\RR$. As in the
construction of static spacetimes, one foliates the interior of
the forward lightcone by hyperboloids $H_T$. By cutting and
inserting strips along  the lifted geodesics on each hyperboloid
$H_T$, one assigns to each cocompact Fuchsian group $\Gamma$ and
each multicurve on $S_\Gamma=\hyp/\Gamma$ a regular domain
$U\in\RR^3$. The cocompact Fuchsian group $\Gamma$ acts on the
domain $U$, and the grafted spacetime of topology $\RR\times S_g$
is obtained by identifying the points in $U$ related by this
action of $\Gamma$.

To give a mathematically precise definition, we follow the
presentation in \cite{bb}. We consider a multicurve on
$S=\hyp/\Gamma$ as in \eqref{multgamma} together with its lift to
a $\Gamma$-invariant weighted multicurve on $\hyp$ as in
\eqref{multlift} and parametrise its geodesics as in \eqref{geods}
\begin{align}
G_I=\{(\bc_{p_i,q_i},w_i)\;|\;i\in I\}.
\end{align}
Furthermore, we choose a basepoint $\bx_0\in\hyp-\bigcup_{i\in I}
\bc_{p_i,q_i}$ that does not lie on any of the geodesics. For each
point $\bx\in \hyp-\bigcup_{i\in I} \bc_{p_i,q_i}$ outside the
geodesics, we choose an arc $\ba_x$ connecting $\bx_0$ and $\bx$,
pointing towards $\bx$ and transverse to each of the geodesics it
intersects. We then define a map $\rho: \hyp-\bigcup_{i\in I}
\bc_{p_i,q_i}\rightarrow \RR^3$ by associating to each
intersection point of $\ba_x$ with one of the geodesics the unit
normal vector of the geodesic pointing towards $\bx$ and
multiplied with the weight
\begin{align}
\label{rhodef2} \rho(\bx)=\!\!\!\!\!\!\!\!\!\!\!\!\!\!\sum_{i\in
I:\; \ba_{x}\cap \bc_{p_i,q_i}\neq\emptyset}
\!\!\!\!\!\!\!\!\!\!\!\! w_i \epsilon_{i,x}
\bn_{p_i,q_i}\qquad\text{for}\;\bx\notin \bigcup_{i\in
I}\bc_{p_i,q_i},
\end{align}
where $\epsilon_{i,x}\in\{\pm 1\}$ is the oriented intersection
number of $\ba_x$ and $\bc_{p_i,q_i}$ with the convention
$\epsilon_{i,x}=1$ if $\bc_{p_i,q_i}$ crosses $\ba_x$ from the
left to the right in the direction of $\ba_x$ and ensures that
$\epsilon_{i,x}w_i\bn_{p_i,q_i}$ points towards $\bx$. Similarly,
for each point $\bx\in \bc_{p_j,q_j}$ that lies on one of the
geodesics, we consider a geodesic ray $\br_x$ starting in $\bx_0$
and through $\bx$, transversal to the geodesics at each
intersection point, and set
\begin{align}
\label{rhogeod}
&\rho_-(\bx)=\!\!\!\!\!\!\!\!\!\!\!\!\!\!\sum_{i\in I-\{j\}:
\;\br_x\cap \bc_{p_i,q_i}\neq\emptyset} \!\!\!\!\!\!\!\!
\!\!\!\!\!\!\!\!w_i \epsilon_{i,x}\bn_{p_i,q_i} & &\rho_+(\bx)=
w_j
\epsilon_{j,x}\bn_{p_jq_j}+\!\!\!\!\!\!\!\!\!\!\!\!\!\!\sum_{i\in
I-\{j\}: \;\br_x\cap \bc_{p_i,q_i}\neq\emptyset} \!\!\!\!\!\!\!\!
\!\!\!\!\!\!\!\! w_i \epsilon_{i,x}\bn_{p_i,q_i}.
\end{align}
On each hyperboloid $H_T$, we now shift the points outside of the
geodesics according to
\begin{align}
\label{pieceshift} T\bx\mapsto
T\bx+\rho(\bx)\qquad\bx\in\hyp-\bigcup_{i\in I}\bc_{p_iq_i}
\end{align}
and replace each geodesic by a strip
\begin{align}
\label{stripshift} T\bx\mapsto
T\bx+t\rho_+(\bx)+(1-t)\rho_-(\bx)\qquad \bx\in \bigcup_{i\in
I}\bc_{p_iq_i}\subset\hyp, t\in[0,1].
\end{align}
From the definitions \eqref{rhodef2}, \eqref{rhogeod} of the maps
$\rho,\rho_\pm$ we see that for each value of $T$, this
corresponds to the grafting procedure for hyperbolic space
described above. The regular domain $U\subset \RR^3$ associated to
the multicurve $G_I$ is the image of the forward lightcone under
this procedure \cite{bb}
\begin{align}
 \label{aconstsurf}&U=\bigcup_{T\in\RR^+} U_T\\
 &U_T=\underbrace{\left\{ T\bx+\rho(\bx)\;|\; \bx \notin
\bigcup_{i\in I}\bc_{p_i,q_i}\right\}}_{=: U_T^0}\cup
\underbrace{\left\{ T\bx+ t\rho_+(\bx)+(1-t)\rho_-(\bx)\;|\;\bx\in
\bigcup_{i\in I}\bc_{p_i,q_i}\,,\;t\in[0,1]\right\}}_{=:
U_T^{G_I}},\nonumber
\end{align} where the two-dimensional surfaces $U_T$ are the images of the hyperboloids
$H_T$, given as a union of shifted pieces $U_T^0$ of hyperboloids
and of strips $U_T^{G_I}$. In particular, the tip of the lightcone
is mapped to the initial singularity $U_0$ of the regular domain
$U$
\begin{align}
U_0=&\left\{ \rho(\bx)\;|\; \bx \notin \bigcup_{i\in
I}\bc_{p_i,q_i}\right\}\cup \left\{
t\rho_+(\bx)+(1-t)\rho_-(\bx)\;|\;\bx\in \bigcup_{i\in
I}\bc_{p_i,q_i}\,,\;t\in[0,1]\right\},
\end{align}
which is a graph (more precisely, a real simplicial tree) with
each vertex corresponding to the area between two geodesics or
between a geodesic and infinity and edges given by $w_i
\epsilon_{i,x}\bn_{p_i,q_i}$.

It is shown in \cite{bg} that the parameter $T$ defines a
cosmological time function $T_c: U\rightarrow \RR^+$
\begin{align}
\label{timedef} T_c(T\bx+\rho(\bx))=T\quad
T_c(T\bx+t\rho_+(\bx)+(1-t)\rho_-(\bx))=T,
\end{align}
and that the surfaces $U_T$ in \eqref{aconstsurf} are surfaces of
constant geodesic distance to the initial singularity $U_0$.

The genus $g$ spacetime associated to the cocompact Fuchsian group
$\Gamma$ and the $\Gamma$-invariant multicurve $G_I$ is then
obtained by identifying on each surface $U_T$ the images of the
points on $H_T$ which are related by the canonical action of
$\Gamma$. This is implemented by defining another  action of the
group $\Gamma$ on $U$. It is shown in \cite{bb} that for
$\Gamma$-invariant multicurves $G_I$ on $\hyp$ the map
\begin{align}
\label{gamgract} &f_{G_I}: \Gamma\rightarrow
\Poi,\;\;f_{G_I}(v)=(v,\rho(\Ad(v)\bx_0))
\end{align}
defines a group homomorphism which leaves each surface $U_T$
invariant, acts on $U$ freely and properly discontinuously and
satisfies
\begin{align}
\label{gaussid} N(f_{G_I}(v)\by)=\Ad(v)N(\by),
\end{align}
where $N: U\rightarrow \hyp$ is the map that associates to each
point in $U$ the corresponding point in $\hyp$
\begin{align}
\label{gauss} N(T\bx+\rho(\bx))=\bx\quad
N(T\bx+t\rho_+(\bx)+(1-t)\rho_-(\bx))=\bx.
\end{align}
The flat (2+1)-spacetime  of genus $g$ associated to the group
$\Gamma$ and the multicurve $G^\Gamma_I$ is defined as the
quotient of $U$ by the action of $\Gamma$ via $f_{G_I}$. Using the
identity \eqref{gaussid}, we find that this amounts to identifying
points $\by,\by'\in \bigcup_{T\in\RR^+} U_T^0$ according to
\begin{align}
\label{finident} & \by\sim\by' \quad\Leftrightarrow\quad\exists
v\in\Gamma:\; N(\by)=\Ad(v)N(\by')\,,\, T_c(\by)=T_c(\by'),
\end{align}
where $T_c: U\rightarrow \RR^+_0$ is the cosmological time
\eqref{timedef}, and for points
$\by,\by'\in\bigcup_{T\in\RR^+}U_T^{G_I}$ parametrised as in
\eqref{aconstsurf}, we have the additional condition $t=t'$.
Hence, two points $\by,\by'\in \bigcup_{T\in\RR^+} U_T^0$ are
identified if and only if they lie on the same surface $U_T$ and
the corresponding points on $\hyp$ are identified by the canonical
action of $\Gamma$ on $\hyp$.

The function $f_{G_I}$ defines the $\Poi$-valued holonomies of the
resulting spacetime. Via the identification $\Gamma\cong
\pi_1(S_g)$ it assigns to each element of the fundamental group
$\pi_1(S_g)$ an element of the group $\Poi$, whose Lorentz
component is the associated element of the Fuchsian group
$\Gamma$. However, in contrast to the static spacetimes considered
above, it is clear from \eqref{gamgract} that in grafted
(2+1)-spacetimes there exist elements of the fundamental group
 whose holonomies have a nontrivial translational
component.

\section{Phase space and Poisson structure in the Chern-Simons formulation of (2+1)-dimensional gravity}
\label{CSbg}

\subsection{The Chern-Simons formulation of (2+1)-dimensional gravity}

The formulation of (2+1)-dimensional gravity as a Chern-Simons
gauge theory is derived from Cartan's description, in which
Einstein's theory of gravity is formulated in terms of a dreibein
of one-forms $e_a$, $a=0,1,2$, and spin connection one-forms
$\omega_a$, $a=0,1,2$, on a spacetime manifold $M$. The dreibein
defines a Lorentz metric $g$ on $M$  via \bea \label{metdreib}
g=\eta^{ab} e_a\otimes e_b, \eea and the one-forms $\omega_a$ are
the coefficients of the spin connection
 \bea
\omega = \omega^a J_a. \eea

To formulate the theory as a Chern-Simons gauge theory, one
combines dreibein and spin-connection into the
 Cartan connection \cite{Sharpe} or
Chern-Simons gauge field \bea \label{Cartan} A = \omega^a J_a +
e_a P^a, \eea  an $iso(2,1)$ valued one form whose curvature
 \bea
\label{decomp} F = T_a  P^a + F_\omega^a\,J_a \eea combines the
curvature and the torsion of the spin connection
\begin{align}
\label{spincurv} &\;F_\omega^a = d\omega^a  + \frac{1}{2}
\epsilon^a_{\;bc} \omega^b\wedge \omega^c &  &T_a = de_a+
\epsilon_{abc} \omega^b e^c.\end{align} This allows one to express
Einstein's equations of motion, the requirements of flatness and
vanishing torsion, as a flatness condition on the Chern-Simons
gauge field \bea \label{flat} F=0.\eea Note, however, that in
order to define a metric $g$ of signature $(1,-1,-1)$ via
\eqref{metdreib},  the dreibein $e$ has to be non-degenerate,
while no such condition is imposed in the corresponding
Chern-Simons gauge theory. It is argued in \cite{Matschull2} for
the case of spacetimes containing particles
 that this leads to
differences in the  global structure of the phase spaces of the
two theories. A further subtlety concerning the phase space in
Einstein's formulation  and the Chern-Simons formulation of
(2+1)-dimensional gravity arises from the presence of large gauge
transformations. It has been shown by Witten \cite{Witten1} that
infinitesimal gauge transformations are on-shell equivalent to
infinitesimal diffeomorphisms in Einstein's formulation of the
theory. This equivalence does not hold for large  gauge
transformations which are not infinitesimally generated and arise
in Chern-Simons theory with non-simply connected gauge groups such
as the group $\Poi$. Nevertheless, configurations related by such
large gauge transformations are identified in the Chern-Simons
formulation of (2+1)-dimensional gravity, potentially causing
further differences in the global structure of the two phase
spaces. However, as we are mainly concerned with the local
properties of the phase space, we will not address these issues
any further in this paper.

In the following we consider spacetimes of topology
$M\approx\RR\times S_g$, where $S_g$ is an orientable two-surface
of genus $g>1$. On such spacetimes, it is possible to give a
Hamiltonian formulation of the theory. One introduces coordinates
$x^0$, $x^1$, $x^2$ on $\RR\times S_g$ such that $x^0$
parametrises $\RR$ and splits the gauge field according to
\begin{align}
\label{gfsplit} A=A_0 dx^0 +A_S,
\end{align}
where $A_S$ is a gauge field on $S_g$ and $A_0: \RR\times
S_\Gamma\rightarrow iso(2,1)$ a  function with values in the
(2+1)-dimensional Poincar\'e algebra. The Chern-Simons action on
$M$ then takes the form \bea \label{CSact}
 S[A_S,A_0]&=&
\int_\RR dx^0\int_{S_g} \tfrac{1}{2}\langle\partial_0 A_S\wedge
A_S\rangle +\langle A_0\,,\, F_S \rangle, \nonumber \eea where
$\langle\,,\,\rangle$ denotes the bilinear form \eqref{inprod} on
$iso(2,1)$, $F_S$ is the curvature of the spatial gauge field
$A_S$ \bea \label{curvv} F_S=d_S A_S + A_S\wedge A_S \eea and
$d_S$ denotes differentiation on the surface $S_g$. The function
$A_0$ plays the role of a Lagrange multiplier, and varying it
leads to the flatness constraint \bea \label{spaceflat} F_S=0,
\eea while variation of $A_S$ yields the evolution equation \bea
 \partial_0 A_S=d_SA_0+[A_S,A_0].\label{asvary}\eea
The action \eqref{CSact} is invariant under gauge transformations
\begin{align} \label{CSgt} &A_0\mapsto \gamma A_0\gamma^\inv+\gamma\partial_0\gamma^\inv\quad
A_S\mapsto\gamma A_S\gamma^\inv+\gamma
d_S\gamma^\inv\quad\text{with}\;\gamma:\RR\times
S_g\rightarrow\Poi,\end{align}  and the phase space of the theory
is the moduli space $\mathcal{M}_g$ of flat $\Poi$-connections
$A_S$ modulo gauge transformations on the spatial surface $S_g$.

\subsection{Holonomies in the Chern-Simons formalism}
\label{holsect}

 Although the moduli space
$\mathcal{M}_g$ of flat $H$-connections is defined as a quotient
of the infinite dimensional space of flat $H$-connections on
$S_g$,  it is of finite dimension $\text{dim}\,
\mathcal{M}_g=2\text{dim}\,H(g-1)$. In the Chern-Simons
formulation of (2+1)-dimensional gravity, we have
$\text{dim}\,\mathcal{M}_g=12(g-1)$, and the finite dimensionality
of $\mathcal{M}_g$ reflects the fact that the theory has no local
gravitational degrees of freedom. From the geometrical viewpoint
this fact can be summarised in the statement that every flat
(2+1)-spacetime is locally Minkowski space. The corresponding
statement in the Chern-Simons formalism is that, due to its
flatness, a gauge field solving the equations of motion can be
trivialised or written as pure gauge on any simply connected
region $R\subset \RR\times S_g$
\begin{align}
\label{triv} A=\gamma d\gamma^\inv=(v^\inv dv,
\Ad(v^\inv)d\bx)\quad\text{with}\;
\gamma^\inv=(v,\bx):R\rightarrow \Poi.
\end{align}
The dreibein on $R$ is then given by $e^a=\Ad(v^\inv)^{ab}dx_b$
and from \eqref{metdreib} it follows that the restriction of the
metric $g$ to $R$ takes the form
\begin{align}
g_{ab}dx^adx^b=(dx^0)^2-(dx^1)^2-(dx^2)^2.
\end{align}
Hence, the translational part of the trivialising function
$\gamma^\inv$ defines an embedding of the region $R$ into
Minkowski space, and the function $\bx(x^0,\cdot)$ gives the
embedding of the surfaces of constant time parameter $x^0$.

\begin{figure}
\vskip .3in \protect\input epsf \protect\epsfxsize=12truecm
\protect\centerline{\epsfbox{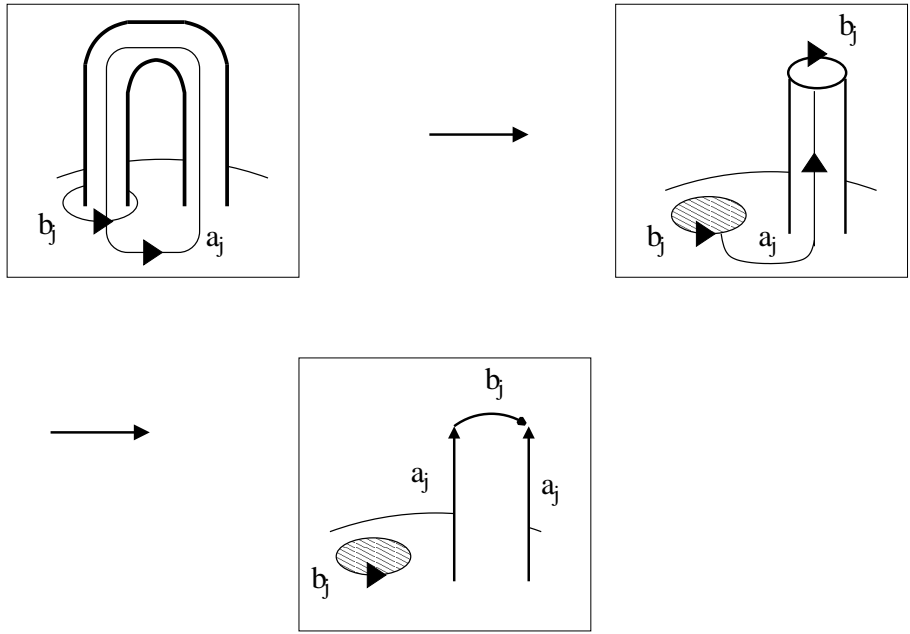}} \caption{Cutting the
surface $S_g$ along the generators of $\pi_1(S_g)$}
\label{cutting}
\end{figure}

\begin{figure}
\vskip .3in \protect\input epsf \protect\epsfxsize=12truecm
\protect\centerline{\epsfbox{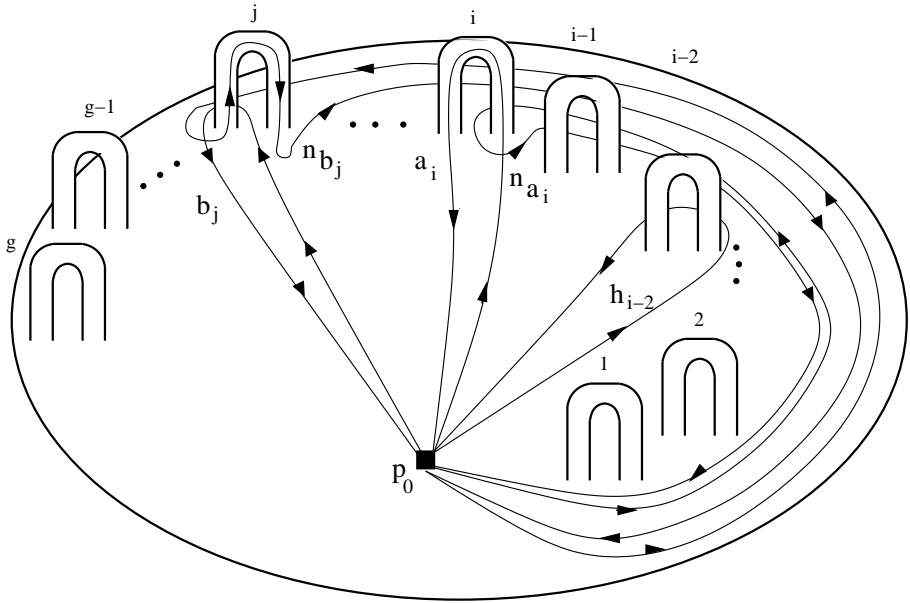}} \caption{Generators and
dual generators of the fundamental group $\pi_1(S_g)$}
\label{pi1gr}
\end{figure}

\begin{figure}
\vskip .3in \protect\input epsf \protect\epsfxsize=12truecm
\protect\centerline{\epsfbox{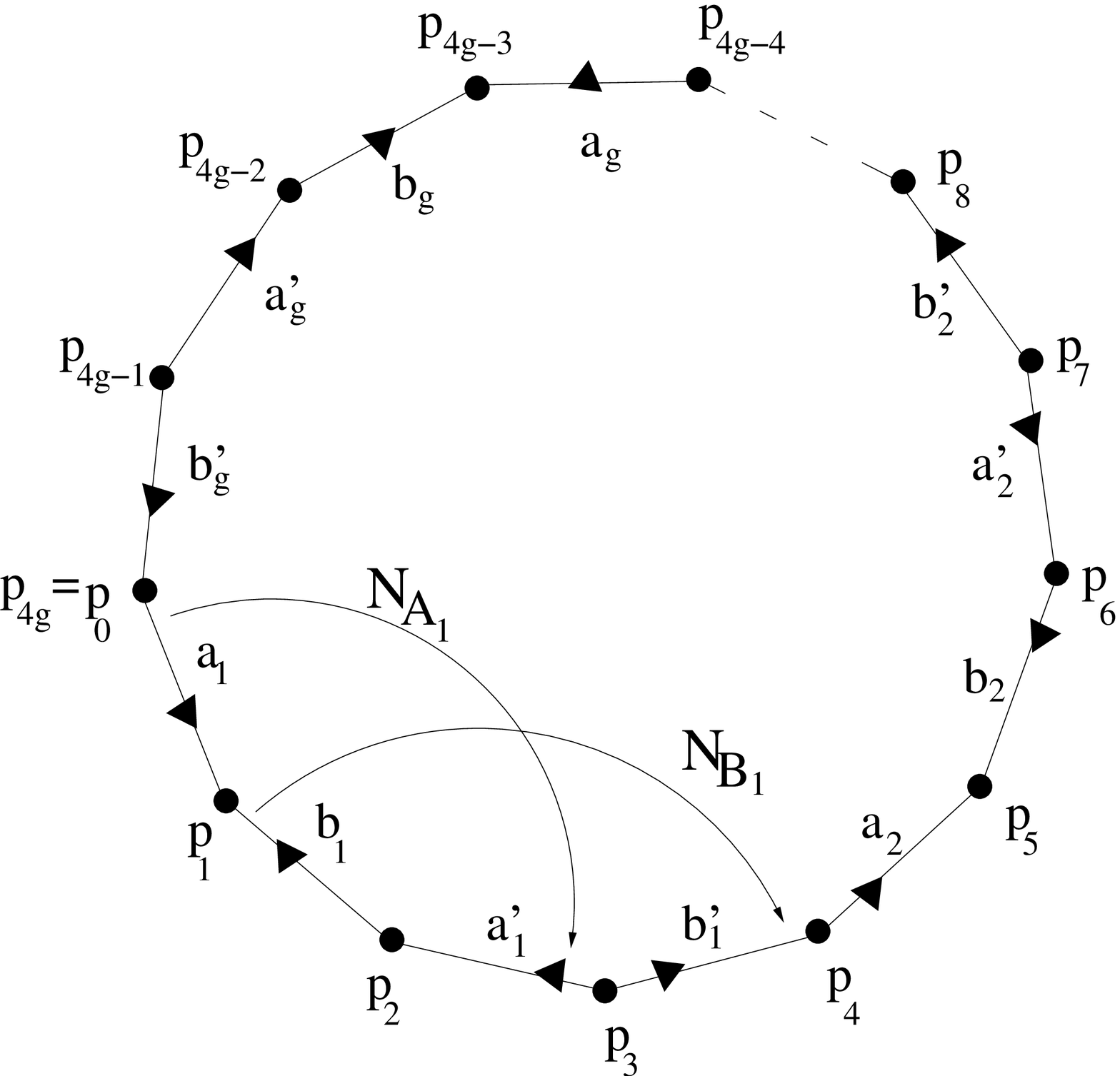}} \caption{The polygon
$P_g$} \label{poly1}
\end{figure}

A maximal simply connected region is obtained by cutting the
spatial surface $S_g$ along a set of generators of the fundamental
group $\pi_1(S_g)$ as  in Fig.\ref{cutting}, which is the approach
pursued by Alekseev and Malkin \cite{AMII}.
The fundamental group of a genus $g$ surface $S_g$ is generated by
two loops $a_i, b_i$, $i=1,\ldots,g$ around each handle, subject
to a single defining relation
\begin{align}
\label{fundgroup} \pi_1(S_g)=\langle a_1,b_1,\ldots, a_g,b_g\,;\,
[b_g,a_g^\inv]\cdots[b_1,a_1^\inv]=1\rangle,
\end{align}
where $[b_i,a_i^\inv]=b_i\circ a_i^\inv\circ b_i^\inv \circ a_i$.
In the
 following we will work with a fixed set of generators
and with a fixed basepoint $p_0$ as shown in Fig.~\ref{pi1gr}.
Cutting  the surface along each generator of the fundamental group
results in a $4g$-gon $P_g$ as pictured in Fig.~\ref{poly1}.
In order to define a gauge field $A_S$ on $S_g$, the function
$\gamma^\inv: P_g\rightarrow \Poi$ must satisfy an overlap
condition relating its values on the two sides corresponding to
each generator of the fundamental group. For any
$y\in\{a_1,b_1,\ldots, a_g,b_g\}$, one must have \cite{AMII}
\begin{align}
\label{csident} & A_S|_{y'}=\gamma d_S\gamma^\inv|_{y'}=\gamma
d_S\gamma^\inv|_{y}=A_S|_y\end{align} which is equivalent to the
existence of a constant Poincar\'e element $N_{Y}=(v_{Y},\bx_{Y})$
such that
\begin{align}
\label{csident2} &\gamma^\inv|_{y'}=N_{Y}
\gamma^\inv|_{y}\quad\text{or, equivalently,} \quad v|_{y'}=v_{Y}
v|_y\qquad \bx|_{y'}=\Ad(v_{Y})\bx|_{y}+\bx_{Y}.
\end{align}
Note that the information about the physical state is encoded
entirely in the Poincar\'e elements $N_X$, $X\in\{A_1,\ldots,
B_g\}$, since transformations of the form $\gamma\mapsto
\tilde\gamma \gamma$ with $\tilde\gamma: S_g\rightarrow\Poi$ are
gauge.
 Conversely, to determine the Poincar\'e elements $N_X$ for a given gauge field, it is not
necessary to know the trivialising function $\gamma$ but only the
embedding of the sides of the polygon $P_g$, which defines them
uniquely via \eqref{csident2}.

We will now relate these Poincar\'e elements to the holonomies  of
our set of generators of the fundamental group $\pi_1(S_g)$. In
the Chern-Simons formalism, the holonomy of a curve
$c:[0,1]\rightarrow S_g$ is given by
\begin{align}
H_c=\gamma(c(1))\gamma^\inv(c(0)),
\end{align}
where $\gamma$ is the trivialising function for the spatial gauge
field $A_S$ on a simply connected region in $S_g$ containing $c$.
By taking the polygon $P_g$ as our simply connected region and
labelling its sides as in Fig.~\ref{poly1}, we find that the
holonomies $A_i$, $B_i$ associated to the curves $a_i$, $b_i$, ,
$i=1,\ldots,g$, are given by \cite{AMII}
\begin{align}
\label{polyhols}
&\ai\!=\!\gamma(p_{4i-3})\gamma(p_{4i-4})^\inv\!\!=\!\gamma(p_{4i-2})\gamma(p_{4i-1})^\inv\;\;\bi\!=\!\gamma(p_{4i-3})\gamma(p_{4i-2})^\inv\!\!\!=\gamma(p_{4i})\gamma(p_{4i-1})^\inv\!\!\!\!.
\end{align}
From the defining relation of the fundamental group, it follows
that they satisfy the relation
\begin{align}
\label{holconst} (u_\infty,
-\Ad(u_\infty)\bj_\infty):=[B_g,A_g^\inv]\cdots[B_1,A_1^\inv]\approx
(1,0).
\end{align}
Using the overlap condition \eqref{csident}, we can express the
value of the trivialising function $\gamma$ at the corners of the
polygon $P_g$ in terms of its value at $p_0$ and find
\begin{align}
\label{cornergf} &\gamma^\inv(p_{4i})= N_{H_i}N_{H_{i-1}}\cdots
N_{H_1}\gamma^\inv(p_0)=\gamma^\inv(p_0)H_1^\inv\cdots
H_{i-1}^\inv H_i^\inv\\
&\gamma^\inv(p_{4i+1})= N_{A_{i+1}}^\inv N_{B_{i+1}}^\inv
N_{A_{i+1}} N_{H_i}\cdots
N_{H_1}\gamma^\inv(p_{0})=\gamma^\inv(p_0)H_1^\inv\cdots
H_{i-1}^\inv H_i^\inv A_{i+1}^\inv
\nonumber\\
&\gamma^\inv(p_{4i+2})=  N_{B_{i+1}}^\inv N_{A_{i+1}}
N_{H_i}\cdots
N_{H_1}\gamma^\inv(p_{0})=\gamma^\inv(p_0)H_1^\inv\cdots
H_{i-1}^\inv H_i^\inv A_{i+1}^\inv B_{i+1}\nonumber\\
&\gamma^\inv(p_{4i+3})=  N_{A_{i+1}} N_{H_i}\cdots
N_{H_1}\gamma^\inv(p_{0})=\gamma^\inv(p_0)H_1^\inv\cdots
H_{i-1}^\inv H_i^\inv A_{i+1}^\inv B_{i+1} A_{i+1},\nonumber
\nonumber
\end{align}
where
\begin{align}
 \label{commnot}
&H_i=(u_{H_i},-\Ad(u_{H_i})\bj_{H_i})=[B_i,A_i^\inv] \qquad
N_{H_i}=(v_{H_i},\bx_{H_i})=[N_\bi, N_\ai^\inv].
\end{align}
Equation \eqref{cornergf} allows us to express the holonomies
$\ai, \bi$ in terms of  $N_\ai,N_\bi$
\begin{align}
\label{holexp3} &A_i= \gamma(p_0)N_{H_1}^\inv\cdots
N_{H_{i-1}}^\inv N_{H_i}^\inv \cdot
N_{\bi}\cdot N_{H_{i-1}} N_{H_{i-2}}\cdots N_{H_1}\gamma^\inv(p_0)\\
&\bi=\gamma(p_0)N_{H_1}^\inv\cdots N_{H_{i-1}}^\inv
N_{H_i}^\inv\cdot  N_{\ai}\cdot N_{H_{i-1}} N_{H_{i-2}}\cdots
N_{H_1}\gamma^\inv(p_0)\nonumber
\end{align}
and by inverting these expressions we obtain
\begin{align}
 \label{holexp4}
&N_{A_i}\!\!=\!\!\gamma^\inv(p_0) H_1^\inv\!\!\!\!\!\cdots\!
H_i^\inv \bi H_{i-1}\!\cdots\! H_1\gamma(p_0)\;\;
N_{B_i}\!\!=\!\!\gamma^\inv(p_0) H_1^\inv\!\!\!\!\!\cdots\!
H_i^\inv \ai H_{i-1}\!\cdots\! H_1\gamma(p_0).
\end{align}
Note that expression \eqref{holexp4} agrees exactly with
\eqref{holexp3} if we exchange $\ai\leftrightarrow N_\ai$,
$\bi\leftrightarrow N_\bi$ and $\gamma^\inv(p_0)\leftrightarrow
\gamma(p_0)$. In particular, up to simultaneous conjugation with
$\gamma^\inv(p_0)$, the Poincar\'e elements $N_\ai,N_\bi$ are the
holonomies along another set of generators of $\pi_1(S_g)$
pictured in Fig.~\ref{pi1gr} and given in terms of the generators
$a_i$, $b_i$ by
\begin{align}
\label{dualgens} &n_{a_i}\!=\! h_1^\inv\circ...\circ h_i^\inv\circ
b_i\circ h_{i-1}\circ...\circ h_1\qquad n_{b_i}\!=\!
h_1^\inv\circ...\circ h_i^\inv\circ a_i\circ h_{i-1}\circ... \circ
h_1,
\end{align}
where $h_i:=[b_i,a_i^\inv]$. From Fig.~\ref{pi1gr}, we see that
this set of generators is dual to the
 set of generators $\{a_1, b_1,\ldots, a_g,b_g\}$ in the
sense that $n_{a_i}$ and $n_{b_i}$, respectively, intersect only
$a_i$ and $b_i$, in a single point.

\subsection{Phase space and Poisson structure}
\label{poisscs}

The description of Chern-Simons theory with gauge group $H$ on
manifolds $\RR\times S_g$ in terms of the holonomies along a set
of generators of the fundamental group $\pi_1(S_g)$ provides an
efficient parametrisation of its phase space $\mathcal{M}_g$.
While the formulation in terms of Chern-Simons gauge fields
exhibits an infinite number of redundant or gauge degrees of
freedom, the characterisation in terms of the holonomies allows
one to describe the moduli space $\mathcal{M}_g$ as a quotient of
a finite dimensional space. It is given as \bea \label{moddef}
\mathcal{M}_g=\{(A_1,B_1,\ldots, A_g,B_g)\in H^{2g}\;|\;
[B_g,A_g^\inv]\cdots [B_1,A_1^\inv]=1\}/H, \eea where the quotient
stands for simultaneous conjugation of all group elements
$\ai,\bi\in H$ by elements of the gauge group $H$. Hence, the
physical observables of the theory are functions on $H^{2g}$ that
are invariant under simultaneous conjugation with $H$ or
conjugation invariant functions of the holonomies associated to
elements of $\pi_1(S_g)$. In the case of the gauge group $\Poi$,
these observables were first investigated in \cite{RN} and
\cite{ahrss}, for the case of disc with punctures representing
massive, spinning particles see also the work of Martin
\cite{martin}, who identifies a complete set of generating
observables and determine their Poisson brackets. In our notation
the two basic observables associated to a general curve
$\eta\in\pi_1(S_g)$ with holonomy $H_\eta=(u_\eta,
-\Ad(u_\eta)\bj_\eta)$, $u_\eta=e^{-p_\eta^a J_a}$, are given by
\begin{align} \label{genmassspin}
&m_\eta^2:=-\bp_\eta^2 & &m_\eta s_\eta:=\bp_\eta\cdot\bj_\eta,
\end{align}
and it  follows directly from the group multiplication law
\eqref{groupmult}, that they are invariant under conjugation of
the holonomies. Furthermore, for a loop $\eta$ around a puncture
representing a massive, spinning particle, $m_\eta$ and $s_\eta$
have the physical interpretation of, respectively, mass and spin
of the particle. In the following we will therefore refer to these
observables as mass and spin of the curve $\eta$.

Although it is possible to determine  the canonical Poisson
brackets of these observables \cite{RN,ahrss}, the resulting
expressions are nonlinear and rather complicated. The main
advantage of the description of the phase space $\mathcal{M}_g$ as
the quotient \eqref{moddef} is that it results in a much simpler
description of the Poisson structure on $\mathcal{M}_g$. Although
the canonical Poisson structure on the space of Chern-Simons gauge
fields does not induce a Poisson structure on the space of
holonomies, it is possible to describe the symplectic structure on
$\mathcal{M}_g$ in terms of an auxiliary Poisson structure on the
manifold $H^{2g}$. The construction is due to Fock and Rosly
\cite{FR} and was developed further by Alekseev, Grosse and
Schomerus \cite{AGSI,AGSII} for the case of Chern-Simons theory
with compact, semisimple gauge groups. A formulation from the
symplectic viewpoint has been derived independently in
\cite{AMII}. In \cite{we1},  this description is adapted to the
universal cover of the (2+1)-dimensional Poincar\'e group and in
\cite{we2} to
 gauge groups of the
form $G\ltimes\mathfrak{g}^*$, where $G$ is a finite dimensional,
connected, simply connected, unimodular Lie group,
$\mathfrak{g}^*$ the dual of its Lie algebra and $G$ acts on
$\mathfrak{g}^*$ in the coadjoint representation. It is shown in
\cite{we1,we2} that in this case, the Poisson structure can be
formulated in terms of a symplectic potential. Although the gauge
group $\Poi=\Lor\ltimes so(2,1)^*$ is not simply connected, the
results of \cite{we1, we2} can nevertheless be applied to this
case\footnote{The assumptions of simply-connectedness and
unimodularity in \cite{we1,we2} are motivated by the absence of
large gauge transformations and by technical simplifications in
the quantisation of the theory but play no role in the classical
results needed in this paper.} and are summarised in the following
theorem.
\begin{theorem} \cite{we2}

\label{symstructth} Consider the Poisson manifold $((\Poi)^{2g},
\Theta)$ with group elements $(A_1,B_1,...,
A_g,B_g)\in(\Poi)^{2g}$ parametrised according to
\begin{align}
\label{holparam} &\ai=(u_\ai,-\Ad(u_\ai)\bj_\ai) & &\bi=(u_\bi,
-\Ad(u_\bi)\bj_\bi) & &i=1,\ldots,g.
\end{align}
and the Poisson structure given by the symplectic form
$\Omega=\delta\Theta$, where
\begin{align}
\label{gravtheta} \Theta=&\sum_{i=1}^g\langle \bj_\ai, \delta(
u_{H_{i-1}}\cdots u_{H_1})( u_{H_{i-1}}\cdots
u_{H_1})^\inv\rangle\\
&\qquad-\langle \bj_\ai, \delta(  u_\ai^\inv   u_\bi^\inv   u_\ai
u_{H_{i-1}}\cdots   u_{H_1})( u_\ai^\inv
  u_\bi^\inv   u_\ai   u_{H_{i-1}}\cdots
u_{H_1})^\inv\rangle\nonumber\\
+&\sum_{i=1}^g\langle \bj_\bi,\delta(  u_\ai^\inv   u_\bi^\inv
u_\ai
  u_{H_{i-1}}\cdots   u_{H_1})(  u_\ai^\inv   u_\bi^\inv   u_\ai u_{H_{i-1}}\cdots
  u_{H_1})^\inv \rangle\nonumber\\
&\qquad-\langle \bj_\bi, \delta(   u_\bi^\inv   u_\ai
u_{H_{i-1}}\cdots
  u_{H_1})(   u_\bi^\inv
  u_\ai   u_{H_{i-1}}\cdots   u_{H_1}
)^\inv \rangle\qquad u_{H_i}=[u_\bi,u_\ai^\inv]\nonumber,
\end{align} and $\delta$ denotes the exterior derivative on
$(\Poi)^{2g}$. Then, the symplectic structure on the moduli space
\begin{align}
\label{moddefpoinc} \mathcal{M}_g=\{(A_1,B_1,\ldots, A_g,B_g)\in
(\Poi)^{2g}\;|\; [B_g,A_g^\inv]\cdots [B_1,A_1^\inv]=1\}/\Poi,
\end{align}
 is obtained from the symplectic form $\Omega=\delta \Theta$ on $(\Poi)^{2g}$ by imposing
 the constraint \eqref{holconst} and dividing by the associated
 gauge transformations which act on the group elements $\ai,\bi$
 by simultaneous conjugation with $\Poi$.
\end{theorem}

\section{Grafting in the Chern-Simons formalism: the transformation of the holonomies}
\label{Csgraft}

In this section we relate the geometrical description of grafted
(2+1)-spacetimes to their description in the Chern-Simons
formalism. We derive explicit expressions for the transformation
of the holonomies $\ai,\bi$, $N_\ai,N_\bi$ of our set of
generators $a_i,b_i\in \pi_1(S_g)$ and their duals
$n_\ai,n_\bi\in\pi_1(S_g)$ under the grafting operation.

We start by considering the static spacetime associated to the
cocompact Fuchsian group $\Gamma$. In this case, we identify the
time parameter $x^0$ in the splitting \eqref{gfsplit} of the gauge
field with the parameter $T$ characterising the hyperboloids
$H_T$. After cutting the spatial surface $S_g$ along our set of
generators  $a_i,b_i\in\pi_1(S_g)$, we obtain the $4g$-gon $P_g$
in Fig.~\ref{poly1} on which the gauge field can be trivialised by
a function
\begin{align}
\label{statgamma}\gamma_{st}^\inv=(v_{st},\bx_{st}):\RR_0^+\times
P_g\rightarrow \Poi \end{align} as in \eqref{triv}. For fixed $T$,
the translational part $\bx_{st}(T,\cdot):\; P_g\rightarrow
P_\Gamma^T$ of $\gamma_{st}^\inv$ maps the polygon $P_g$ to the
polygon $P_\Gamma^T\subset H_T$ defined by the Fuchsian group
$\Gamma$ , such that the images of sides and corners of $P_g$ are
the corresponding sides and corners of $P_\Gamma^T$.

 By choosing coordinates on $P_g$, it
is in principle possible to give an explicit expression for the
trivialising function $\gamma^\inv_{st}: \RR^+_0\times
P_g\rightarrow \Poi$. However, in order to determine the
holonomies $\ai,\bi$ and  $N_\ai,N_\bi$, it is sufficient to know
the embedding of the sides and corners of $P_g$. As the two sides
of the polygon $P_\Gamma^T$ corresponding to each generator
$a_i,b_i\in\pi_1(S_g)$ are mapped into each other by the
generators of $\Gamma$ according to \eqref{fuchsels}, the overlap
condition \eqref{csident2} for the trivialising function
$\gamma^\inv_{st}$ becomes
\begin{align}
\label{stativov} v_{st}(T,\cdot)|_{y'}=v_Y
v_{st}(T,\cdot)|_{y}\qquad\bx_{st}(T,\cdot)|_{y'}=\Ad(v_Y)\bx_{st}(T,\cdot)|_{y},
\end{align}
where $y\in\{a_1,\ldots,b_g\}$, $Y\in\{A_1,\ldots, B_g\}$ and
$v_Y$ denotes the  associated generator of $\Gamma$. The
holonomies $N_\ai,N_\bi$ are therefore given by
\begin{align}
\label{statichols} N_X=(v_X,0)\qquad X\in\{A_1,\ldots, B_g\}.
\end{align}
Their translational components vanish, and the same holds for the
holonomies $\ai,\bi$ up to conjugation with the Poincar\'e element
$\gamma(p_0)$ associated to the basepoint.

We  now consider the (2+1)-spacetimes obtained from the static
spacetime associated to $\Gamma$ via grafting along a closed,
simple geodesic $\lambda\in\pi_1(S_\Gamma)$ on $S_\Gamma$ with
weight $w$. As discussed in Sect.~\ref{hypgeom}, this geodesic
lifts to a $\Gamma$-invariant multicurve on $\hyp$
\begin{align}
\label{geodfamily} G=\{(\Ad(v)\bc_{p,q}, w)\;|\; v\in\Gamma\},
\end{align}
where $\bc_{p,q}$ is the lift of $\lambda$, parametrised as in
\eqref{geods} with $\bp\in P_\Gamma^1$. As the geodesic
$\bc_{p,q}$ is the lift of a simple closed geodesic on $S_\Gamma$,
there exists a nontrivial element $\tilde v=e^{\alpha n_{p,q}^a
J_a} \in\Gamma$ with $\alpha\in\RR^+$, the holonomy of $\lambda$
defined up to conjugation, that maps the geodesic $\bc_{p,q}$ to
itself.
 More precisely, the
geodesic $\bc_{p,q}$ traverses a sequence of polygons
\begin{align}
P_1\!=\!P_\Gamma^1,\;\;\; P_2\!=\!\Ad(v_{r})P_\Gamma^1,\;\;\;
P_3\!=\!\Ad(v_{r-1} v_r)P_\Gamma^1,\ldots,
P_{r+1}\!=\!\Ad(v_1\cdots v_r)P_\Gamma^1\!=\!\Ad(\tilde
v)P_\Gamma^1
\end{align}
mapped into each other by group elements $v_i\in\Gamma$, until it
 reaches a point $\bp'=\Ad(v_1\cdots v_r)\bp=\Ad(\tilde v)\bp\in
 P_{r+1}$ identified with $\bp$. In particular, this
 implies that the geodesics in \eqref{geodfamily} do not have
 intersection points with the corners of the polygons in the
 tesselation of $H_T$. In the following we therefore take the
 corner $\bx_{st}(T, p_0)$ as our basepoint $\bx_0$ and parametrise
 \begin{align}
 \label{bsptpar}
 \gamma^\inv_{st}(T, p_0)=(v_0, \bx_0).
 \end{align}

As each generator of the Fuchsian group $\Gamma$ maps the polygon
$P_\Gamma^T$ into one of its neighbours, we can express the group
elements $v_i\in\Gamma$ in terms of the generators and their
inverses as
\begin{align}&v_k= v_{X_r}^{\alpha_r}\cdots
v_{X_{k+1}}^{\alpha_{k+1}} v_{X_{k}}^{\alpha_k}
v_{X_{k+1}}^{-\alpha_{k+1}}\cdots v_{X_r}^{-\alpha_r} &
\label{polyidentpar}&\tilde v=e^{\alpha n_{p,q}^a J_a}=v_1\cdots
v_r=v_{X_r}^{\alpha_r}\cdots v_{X_1}^{\alpha_1}
\end{align}
with $v_{X_i}\in\{v_{A_1},\ldots v_{B_g}\}$, $\alpha_i=\pm 1$. To
determine the  map $\rho$ in \eqref{rhodef2} for the grafting
along the multicurve \eqref{geodfamily}, we note that a general
geodesic $\Ad(v)\bc_{p,q}=\bc_{\Ad(v)p,\Ad(v)q}$, $v\in\Gamma$,
is mapped to itself by the element $v\tilde v v^\inv\in\Gamma$ and
has the unit normal vector
$\bn_{\Ad(v)p,\Ad(v)q}=\Ad(v)\bn_{p,q}$. The map $\rho$ in
\eqref{rhodef2} is therefore given by
\begin{align}
\label{rhopolydef}
\rho(\bx)=w\!\!\!\!\!\!\!\!\!\!\!\!\!\!\!\!\!\!\!\!\sum_{v\in\Gamma:\;
\ba_{x}\cap \bc_{\Ad(v)p,\Ad(v)q}\neq\emptyset}
\!\!\!\!\!\!\!\!\!\!\!\!\!\!\!\!\!\!\!\!\epsilon_{v,x}
\Ad(v)\bn_{p,q}\qquad\text{for}\;\bx\notin \bigcup_{v\in\Gamma}
\bc_{\Ad(v)p,\Ad(v)q}
\end{align}
and \eqref{rhogeod} implies for points $\bx\in
\bc_{\Ad(v_x)p,\Ad(v_x)q}$, $v_x\in\Gamma$, on one of the
geodesics
\begin{align}
\label{rhopolydef2}
\rho_-(\bx)=w\!\!\!\!\!\!\!\!\!\!\!\!\!\!\!\!\!\!\!\!\!\!\!\!\!\!\sum_{v\in\Gamma-\{v_x\}:\;
\br_{x}\cap
\bc_{\Ad(v)p,\Ad(v)q}\neq\emptyset}\!\!\!\!\!\!\!\!\!\!\!\!\!\!\!\!\!\!\!\!\!\!\!\!\!\!\!
\epsilon_{v,x}\Ad(v)\bn_{p,q}\qquad\rho_+(\bx)=\rho_-(\bx)+w\epsilon_{v_x,x}\,\Ad(v_x)\bn_{p,q}.
\end{align}
We are now ready to determine the transformation of the holonomies
$\ai,\bi$ and  $N_\ai,N_\bi$ under  grafting along $\lambda$. We
identify the time parameter $x^0$ in \eqref{gfsplit} with the
cosmological time $T$ of the regular domain \eqref{aconstsurf}
associated to the multicurve \eqref{geodfamily}. For fixed $T$,
the translational part of the trivialising function
$\gamma^\inv=(v,\bx): \RR^+_0\times P_g\rightarrow \Poi$ maps the
polygon $P_g$ to the image $P_{\Gamma,G}^T\subset U_T$ of
$P_\Gamma^T$ under the grafting operation
\begin{align}
\label{embedgr} &\bx(T,\cdot): P_g\rightarrow P_{\Gamma,G}^T\\
&P_{\Gamma,G}^T=\left\{ T\bx+\rho(\bx)\;|\; \bx \in P_\Gamma^1-
\bigcup_{v\in \Gamma}\bc_{\Ad(v)p,\Ad(v)q}\right\}\nonumber\\
&\qquad\cup \left\{ T\bx+ t\rho_+(\bx)+(1-t)\rho_-(\bx)\;|\;\bx\in
P_\Gamma^1\cap \bigcup_{v\in
\Gamma}\bc_{\Ad(v)p,\Ad(v)q}\,,\;t\in[0,1]\right\}\subset
U_T\nonumber
\end{align}
Again, we do not need an explicit expression for the embedding
function $\gamma^\inv$ but can determine the holonomies $\ai,\bi$
and  $N_\ai,N_\bi$ from the embedding of the sides of the polygon
$P_g$. For this,  we consider a side $y\in\{a_1,b_1\ldots,
a_g,b_g\}$  of the polygon $P_g$ and denote by $q_i^Y$, $q_f^Y$,
respectively,  its starting and endpoint. In the case of the
static spacetime associated to $\Gamma$, the  holonomy $Y_{st}$
along $y$ with respect to the basepoint $p_0$ is given by
\begin{align}
Y_{st}=\gamma_{st}(T,q^Y_f)\gamma^\inv_{st}(T,q_i^Y).
\end{align}
Since the geodesics in \eqref{geodfamily} do not intersect the
corners of the polygon, the embedding of starting and endpoint of
$y$ in the resulting regular domain  is
\begin{align}
\label{shift0} &\bx(T,q_i^Y)=\bx_{st}(T,q_i^Y)+\rho(q_i^Y) &
&\bx(T,q_f^Y)=\bx_{st}(T,q_f^Y)+\rho(q_f^Y),
\end{align}
where here and in the following $\rho(q)$, $q\in P_g$, stands for
$\rho(\bx_{st}(1,q))$. This implies
\begin{align}
\label{shift1} &\gamma^\inv(T,q_{i,f}^Y)=
(v_{st}(T,q_{i,f}^Y),\bx_{st}(T,q_{i,f}^Y)+\rho(q_{i,f}^Y))=(1,
\rho(q_{i,f}^Y))\cdot\gamma^\inv_{st}(T,q_{i,f}^Y),
\end{align}
 and the holonomy $Y$ becomes
\begin{align}
Y=Y_{st}\cdot (1,
-\Ad(v_{st}^\inv(T,q_i^Y))(\rho(q_f^Y)-\rho(q_i^Y)))\nonumber.
\end{align} From expression \eqref{rhopolydef}
for the map $\rho$ we deduce
\begin{align}
\rho(q_f^Y)-\rho(q^Y_i)=
w\!\!\!\!\!\!\!\!\!\!\!\!\!\!\!\sum_{v\in\Gamma: y\cap
\bc_{\Ad(v)p,\Ad(v)q}\neq\emptyset}
 \!\!\!\!\!\!\!\!\!\!\!\!\!\!\!\epsilon_{v,y} \Ad(v)\bn_{p,q},
\end{align}
where $\epsilon_{v,y}$ is the oriented intersection number of
$\bc_{\Ad(v)p,\Ad(v)q}$ and $y$, taken to be positive if
$\bc_{\Ad(v)p,\Ad(v)q}$ crosses $y$ from the left to the right in
the direction of $y$. In order to determine the transformations of
the holonomies $\ai,\bi$, we therefore need to determine the
intersection points of the multicurve \eqref{geodfamily} with the
sides of the polygon $P_\Gamma^1$ and the oriented intersection
numbers $\epsilon_{v,y}$.

As the geodesic $\bc_{p,q}$ intersects the side
$\Ad(v_{r-k+2}\cdots v_r)x \subset P_k$ of the polygon $P_k$ if
and only if the geodesic $\Ad(v_r^\inv\cdots v_{r-k+2}^\inv)
\bc_{p,q}=\Ad(v_{X_{r-k+1}}^{\alpha_{r-k+1}}\cdots
v_{X_1}^{\alpha_1})\bc_{p,q}$ intersects the side $x\subset
P_\Gamma^1$, the geodesics in \eqref{geodfamily} which have
intersections points with the sides of  $P_\Gamma^1$ are
\begin{align}
\label{cdef}
\bc_1=\bc_{p,q},\,\;\bc_2=\Ad(v_{X_1}^{\alpha_1})\bc_{p,q},\,\;
\bc_3=\Ad(v_{X_2}^{\alpha_2}v_{X_1}^{\alpha_1})\bc_{p,q},\;\ldots,\;
\bc_{r}=\Ad(v_{X_{r-1}}^{\alpha_{r-1}}\cdots
v_{X_1}^{\alpha_1})\bc_{p,q}.
\end{align}
The intersections of the multicurve \eqref{geodfamily} with a
given side $y\subset P^1_\Gamma$ are in one-to-one correspondence
with factors $v_{X_k}^{\alpha_k}$, $X_k=Y$, in
\eqref{polyidentpar}, and the geodesic in \eqref{cdef}
intersecting $y$ is $\bc_k$ if $\alpha_k=1$ and $\bc_{k+1}$ if
$\alpha_k=-1$. Similarly, intersections with the side $y'$ are
also in one-to-one correspondence with factors
$v_{X_k}^{\alpha_k}$, $X_k=Y$, but the intersection takes place
with $\bc_k$ for $\alpha_k=-1$ and with $\bc_{k+1}$ for
$\alpha_k=1$. Taking into account the orientation of the sides
$a_i,b_i, a'_i, b'_i$ in the polygon $P_g$, see Fig.~\ref{poly1},
we find that intersections with sides $a_i$ and $a'_i$ have
positive intersection number for $\alpha_k=1$ and negative
intersection number for $\alpha_k=-1$, while the intersection
numbers for sides $b_i, b'_i$ are positive and negative,
respectively, for $\alpha_k=-1$ and $\alpha_k=1$. Hence, we find
that the transformation of the holonomy $Y=(u_Y,-\Ad(u_Y)\bj_Y)$
under grafting along $\lambda$ is given by
\begin{align}
\label{grafttrafo}
&u_Y\mapsto u_Y\\
&\bj_Y\mapsto\bj_Y +\epsilon_Y
w\;\Ad(v_{st}^\inv(q^Y_i))\left(\!\!\!\!\!\sum_{\quad i:
X_i=Y,\alpha_i=1}\!\!\!\!\!\Ad(v_{X_{i-1}}^{\alpha_{i-1}}\cdots
v_{X_1}^{\alpha_1})\bn_{p,q} - \!\!\!\!\! \sum_{i:
X_i=Y,\alpha_i=-1}\!\!\!\!\!\Ad(v_{X_{i}}^{\alpha_{i}}\cdots
v_{X_1}^{\alpha_1})\bn_{p,q}\right),\nonumber
\end{align}
where the overall sign $\epsilon_Y$ is positive for $Y=\ai$ and
negative for $Y=\bi$. Note that \eqref{grafttrafo} is invariant
under conjugation
 of the group element $\tilde
v=v_{X_r}^{\alpha_r}\cdots v_{X_1}^{\alpha_1}\in\Gamma$ associated
to the geodesic $\bc_{p,q}$ with elements of  $\Gamma$. Although
such a conjugation can give rise to additional intersection
points, the
 identity $\Ad( v_{X_r}^{\alpha_r}\cdots
v_{X_1}^{\alpha_1})\bn_{p,q}=\bn_{p,q}$ implies that their
contributions to the transformation \eqref{gencurvegraft} cancel.
Hence, the transformation \eqref{grafttrafo} depends only on the
geodesic $\lambda\in\pi_1(S_\Gamma)$ and not on the choice of the
lift $\bc_{p,q}$.

To deduce the transformation of the holonomies $\ai,\bi$, we
determine the corresponding starting and endpoints from
Fig.~\ref{poly1}. For $Y=A_i$, starting and end point are given by
$q^\ai_i=p_{4(i-1)}$, $q^\ai_f=p_{4i-3}$, for $Y=B_i$ by
$q^\bi_i=p_{4i-2}$, $q^\bi_f=p_{4i-3}$, and \eqref{cornergf}
implies
\begin{align}
\label{fact} &v_{st}^\inv(q_i^\ai)=v_0^\inv v_{H_{1}}^\inv\cdots
v_{H_{i-1}}^\inv & &v_{st}^\inv(q_i^\bi)=v_0^\inv
v_{H_{1}}^\inv\cdots v_{H_{i-1}}^\inv v_\ai^\inv v_\bi.
\end{align}
Taking into account the oriented intersection numbers, we find
that the holonomies $\ai,\bi$ transform under grafting along
$\lambda$ according to
\begin{align}
\label{graftaction}  &\bj_\ai\!\!\mapsto\!
\bj_\ai\!\!\!+\!\!w\Ad(v_0^\inv v_{H_1}^\inv\!\cdots\!
v^\inv_{H_{i-1}})\!\!\left(\!\!\!\!\!\!\!\!\!\!\!\!\!\!\!\!\!\!\sum_{\quad\qquad
k:X_k=\ai,\alpha_k=1}\!\!\!\!\!\!\!\!\!\!\!\!\!\!\!\!\!\!\!\Ad(v^{\alpha_{k-1}}_{X_{k-1}}\cdots
v^{\alpha_1}_{X_1})\bn_{p,q}\!\!-\!\!\!\!\!\!\!\!\!\!\!\!\!\!\sum_{k:
X_k=\ai,\alpha_k=-1}\!\!\!\!\!\!\!\!\!\!\!\!\Ad(v^{\alpha_{k}}_{X_{k}}\cdots
v^{\alpha_1}_{X_1})\bn_{p,q}\!\!\right)\\
 &\bj_\bi\!\!\mapsto\! \bj_\bi\!\!-\!\! w\Ad(v_0^\inv
v_{H_1}^\inv\!\cdots\! v^\inv_{H_{i-1}}v_\ai^\inv
v_\bi)\!\!\left(\!\!\!\!\!\!\!\!\!\!\!\!\!\!\!\!\!\!\sum_{\quad\qquad
k:X_k=\bi,\alpha_k=1}\!\!\!\!\!\!\!\!\!\!\!\!\!\!\!\!\!\!\!\!\Ad(v^{\alpha_{k-1}}_{X_{k-1}}\cdots
v^{\alpha_1}_{X_1})
\bn_{p,q}\!\!-\!\!\!\!\!\!\!\!\!\!\!\!\!\!\sum_{k:
X_k=\bi,\alpha_k=-1}\!\!\!\!\!\!\!\!\!\!\!\!\Ad(v^{\alpha_{k}}_{X_{k}}\cdots
v^{\alpha_1}_{X_1}) \bn_{p,q}\!\!\right)\nonumber.
\end{align}
Equivalently, we could have determined the transformation of the
holonomies from the sides $a'_i,b'_i$. In this case,
$q^Y_i=p_{4i-1}$ for both $y=a'_i, b'_i$ and therefore
$v_{st}^\inv(q_i^Y)=v_0^\inv v_{H_{1}}^\inv\cdots v_{H_{i-1}}^\inv
v_\ai^\inv= v_{st}^\inv(q_i^\bi)v_\bi^\inv= v_{st}^\inv(q_i^\ai)
v_\ai^\inv$, which together with the remark before
\eqref{grafttrafo} yields the same result.

With the interpretation of the holonomies $\ai,\bi$ as the
different factors in the product $(\Poi)^{2g}$,
\eqref{graftaction} defines a map
 $Gr_{w \lambda}:(\Poi)^{2g}\rightarrow(\Poi)^{2g}$
which leaves the submanifold $(\Lor)^{2g}\subset(\Poi)^{2g}$
invariant. The transformation of the holonomy of a general curve
$\eta=y_s^{\beta_s}\circ\ldots\circ
y_1^{\beta_1}\in\pi_1(p_0,S_g)$ under $Gr_{w\lambda}$ is then
obtained by writing the curve as a product in the generators
$\ai,\bi$. Parametrising the associated holonomy as
$H_\eta=(u_\eta, -\Ad(u_\eta)\bj_\eta)$ as in \eqref{gparam}, we
find that the vector $\bj_\eta$ is given by
\begin{align}
\label{etavect} &\bj_\eta=\sum_{i=1, \beta_i=1}^s
\Ad(u_{Y_1}^{-\beta_1}\cdots
u_{Y_{i-1}}^{-\beta_{i-1}})\bj_{Y_i}-\sum_{i=1, \beta_i=-1}^r
\Ad(u_{Y_1}^{-\beta_1}\cdots u_{Y_{i}}^{-\beta_{i}})\bj_{Y_i}
\end{align}
and using \eqref{graftaction}, we obtain the following theorem.

\begin{theorem}\label{graftsummary} For  $\eta=y_s^{\beta_s}\circ\ldots\circ
y_1^{\beta_1}\in\pi_1(S_g)$ with $y_i\in\{a_1,\ldots,b_g\}$,
$\beta_i\in\{\pm 1\}$,  the transformation of the associated
holonomy under grafting along $\lambda$ is given by
\begin{align}
\label{gencurvegraft} Gr_{w\lambda}&:\;u_\eta\mapsto u_\eta\\
 \bj_\eta\mapsto
&\bj_\eta +w\sum_{i=1}^g
\left(\sum_{Y_j=\ai,\beta_j=1}\Ad(u_{Y_1}^{-\beta_1}\cdots
u_{Y_{j-1}}^{-\beta_{j-1}})-\!\!\!\!\!\!\!\!\sum_{Y_j=\ai,
\beta_j=-1} \!\!\!\!\!\!\!\!\Ad(u_{Y_1}^{-\beta_1}\cdots
u_{Y_{j}}^{-\beta_{j}})\right)\cdot\nonumber\\
&\cdot\Ad(v_0^\inv v^\inv_{H_{1}} \cdots
v^\inv_{H_{i-1}})\left(\sum_{k:X_k=\ai,\alpha_k=1}\!\!\!\!\!\!\!\!\!\!\!\!\Ad(v^{\alpha_{k-1}}_{X_{k-1}}\cdots
v^{\alpha_1}_{X_1})\bn_{p,q}-\!\!\!\!\!\!\!\!\!\!\!\!\sum_{k:
X_k=\ai,\alpha_k=-1}\!\!\!\!\!\!\!\!\!\!\!\!\Ad(v^{\alpha_{k}}_{X_{k}}\cdots
v^{\alpha_1}_{X_1})\bn_{p,q}\right)\nonumber\\
&-w\sum_{i=1}^g
\left(\sum_{Y_j=\bi,\beta_j=1}\Ad(u_{Y_1}^{-\beta_1}\cdots
u_{Y_{j-1}}^{-\beta_{j-1}})-\!\!\!\!\!\!\!\!\sum_{Y_j=\bi,
\beta_j=-1} \!\!\!\!\!\!\!\!\Ad(u_{Y_1}^{-\beta_1}\cdots
u_{Y_{j}}^{-\beta_{j}})\right)\cdot\nonumber\\
&\cdot\Ad(v_0^\inv v_{H_1}^\inv \cdots v_{H_{i-1}}^\inv v_\ai^\inv
v_\bi)
\left(\sum_{k:X_k=\bi,\alpha_k=1}\!\!\!\!\!\!\!\!\!\!\!\!\Ad(v^{\alpha_{k-1}}_{X_{k-1}}\cdots
v^{\alpha_1}_{X_1})\bn_{p,q}-\!\!\!\!\!\!\!\!\!\!\!\!\sum_{k:
X_k=\bi,\alpha_k=-1}\!\!\!\!\!\!\!\!\!\!\!\!\Ad(v^{\alpha_{k}}_{X_{k}}\cdots
v^{\alpha_1}_{X_1})\bn_{p,q}\right).\nonumber
\end{align}
\end{theorem}
Although the formula  for the transformation of $\bj_\eta$ appears
rather complicated, one can give a heuristic  interpretation of
the various factors \eqref{gencurvegraft}. For this recall that,
up to conjugation, the group element $\tilde v\in\Gamma$ gives the
holonomy of the geodesic $\lambda$ and consider the associated
element $\lambda=n_{X_r}^{\alpha_r}\circ\ldots\circ
n_{X_1}^{\alpha_1}\in\pi_1(p_0,S_g)$ of the fundamental group
based at $p_0$.  The holonomy along this element is
\begin{align}
\label{holgrcurve}
H_{\lambda}=(u_\lambda,-\Ad(u_\lambda)\bj_\lambda)=\gamma(p_0)N_{X_r}^{\alpha_r}\cdots
N_{X_1}^{\alpha_1}\gamma^\inv(p_0)\qquad u_\lambda=e^{-p_\lambda^a
J_a},
\end{align}
and from \eqref{polyidentpar} it follows that the unit vector
$\bn_{p,q}$ is given by
\begin{align}
\label{nprel}
\bn_{p,q}=-\Ad(v_0)\hat\bp_{\lambda}=-\tfrac{1}{m_\lambda}\Ad(v_0)\bp_{\lambda}.
\end{align}
Hence, the terms $\Ad(v^{\alpha_{k-1}}_{X_{k-1}}\cdots
v^{\alpha_1}_{X_1})\bn_{p,q}$, and
$\Ad(v^{\alpha_{k}}_{X_{k}}\cdots v^{\alpha_1}_{X_1})\bn_{p,q}$ in
\eqref{gencurvegraft} can be viewed as the parallel transport
along $\lambda$ of the vector $\hat\bp_{\lambda}$ from the
starting point of $\lambda$ to the intersection point with the
sides $a_i,b_i$ of the polygon $P_g$. The terms $\Ad(v_0^\inv
v^\inv_{H_{1}} \cdots v^\inv_{H_{i-1}})$ and $\Ad(v_0^\inv
v_{H_1}^\inv \cdots v_{H_{i-1}}^\inv v_\ai^\inv v_\bi)$ transport
the vector from the point $p_0$ to the starting point of,
respectively, sides $\ai$ and $\bi$ of $P_g$. Finally, the terms
$\Ad(u_{Y_1}^{-\beta_1}\cdots u_{Y_{j-1}}^{-\beta_{j-1}})$ and
$\Ad(u_{Y_1}^{-\beta_1}\cdots u_{Y_{j}}^{-\beta_{j}})$ describe
the parallel transport along the curve $\eta$ from its
intersection point with $\lambda$ to its starting point $p_0$. We
will give a more detailed and precise interpretation of this
formula in Sect.~\ref{grdtrel}, where we discuss the link between
grafting and Dehn twists.


\section{Grafting and Poisson structure}
\label{massgen}

In this section, we give explicit expressions for Hamiltonians on
the Poisson manifold $((\Poi)^{2g},\Theta)$ which generate the
transformation \eqref{gencurvegraft} of the holonomies under
grafting via the Poisson bracket. As
 we have seen in Sect.~\ref{Csgraft} that the grafting operation
is most easily described by parametrising one of the holonomies in
question in terms of  $\ai,\bi$ and the other one in terms of
$N_\ai,N_\bi$, the first step  is to derive an expression for the
symplectic potential \eqref{gravtheta}  involving the components
of both $\ai,\bi$ and $N_\ai,N_\bi$.  We then  prove that the
transformation \eqref{gencurvegraft} of the holonomies under
grafting along a geodesic  $\lambda\in\pi_1(S_g)$ with weight $w$
is generated by $w m_\lambda$, where $m_\lambda$ is the mass of
$\lambda$ defined as in \eqref{genmassspin}. Finally, we use this
result to investigate the properties of the grafting
transformation $Gr_{w\lambda}:(\Poi)^{2g}\rightarrow (\Poi)^{2g}$
and prove a  relation between the Poisson brackets of mass and
spin for general elements $\lambda,\eta\in\pi_1(S_g)$.

\subsection{The Poisson structure in terms  of the dual generators}
\label{dualpoiss}

In order to derive an expression for the symplectic potential
\eqref{gravtheta} in terms of  both $\ai,\bi$ and $N_\ai,N_\bi$,
we need to express the Lorentz and translational components of the
holonomies $\ai,\bi$ and $N_\ai,N_\bi$ in terms of each other via
\eqref{holexp3} and \eqref{holexp4}. For the Lorentz components,
we can simply replace $\ai,\bi$ with $u_\ai,u_\bi$, $N_\ai,N_\bi$
with $v_\ai,v_\bi$ and $\gamma^\inv(p_0)$ with $v_0$ in
\eqref{holexp3}, \eqref{holexp4} and obtain
\begin{align}
\label{holexplor} &v_{A_i}=v_0 u_{H_1}^\inv\cdots u_{H_i}^\inv
\cdot u_\bi\cdot u_{H_{i-1}}\cdots u_{H_1}v_0^\inv & &v_{B_i}=v_0
u_{H_1}^\inv\cdots u_{H_i}^\inv \cdot u_\ai\cdot u_{H_{i-1}}\cdots
u_{H_1}v_0^\inv\\
&u_{A_i}= v_0^\inv v_{H_1}^\inv\cdots v_{H_i}^\inv \cdot
v_{\bi}\cdot v_{H_{i-1}}\cdots v_{H_1}v_0 &
 &u_{B_i}= v_0^\inv v_{H_1}^\inv\cdots v_{H_i}^\inv
\cdot v_{\ai}\cdot v_{H_{i-1}}\cdots v_{H_1}v_0,\nonumber
\end{align}
where $u_{H_i}=[u_\bi, u_\ai^\inv]$, $v_{H_i}=[v_\bi,v_\ai^\inv]$.
The corresponding expressions for the translational components
require some computation. Inserting the parametrisation of the
holonomies $\ai,\bi$ into \eqref{holexp4} and using
\eqref{holexplor}, we find
\begin{align}
\label{xjexpr} \bx_\ai=&(1-\Ad(v_\ai))\left(\bx_0+\sum_{k=1}^{i-1}
(1-\Ad(v_{A_k}))\bl_{A_k}+(1-\Ad(v_{B_k}))\bl_{B_k}\right)\\
&+\Ad(v_\bi)\bl_\bi+(1-\Ad(v_\ai))\bl_\ai+(1-\Ad(v_\bi))\bl_\bi\nonumber\\
\bx_\bi=&(1-\Ad(v_\bi))\left(\bx_0+\sum_{k=1}^{i-1}
(1-\Ad(v_{A_k}))\bl_{A_k}+(1-\Ad(v_{B_k}))\bl_{B_k}\right)\nonumber\\
&-\Ad(v_\bi)\bl_\ai+(1-\Ad(v_\ai))\bl_\ai+(1-\Ad(v_\bi))\bl_\bi\nonumber\\
\nonumber\\
 \label{goodjs} \bl_\ai=&\Ad(v_{H_{i-1}}\cdots
v_{H_1}v_0)\bj_\ai=\Ad(v_0u_{H_1}^\inv\cdots u_{H_{i-1}}^\inv u_\ai^\inv) \cdot \Ad(u_\ai)\bj_\ai\\
\bl_\bi=&-\Ad(v_\bi^\inv v_\ai v_{H_{i-1}}\cdots
v_{H_1}v_0)\bj_\bi=-\Ad(v_0u_{H_1}^\inv\cdots u_{H_{i-1}}^\inv
u_\ai^\inv) \cdot \Ad(u_\bi)\bj_\bi\nonumber,
\end{align}
and an analogous calculation for \eqref{holexp3} yields
\begin{align}
\label{jxpr}
\bj_\ai=&-(1-\Ad(u_\ai^\inv))\left(\Ad(v_0^\inv)\bx_0+\sum_{k=1}^{i-1}(1-\Ad(u_{A_k}))\bfh_{A_k}+(1-\Ad(u_{B_k}))\bfh_{B_k}\right)\\
&+\Ad(u_\ai^\inv)\left((1-\Ad(u_\ai))\bfh_\ai+(1-\Ad(u_\bi))\bfh_\bi+\Ad(
u_\bi)\bfh_\bi\right)\nonumber\\
\bj_\bi=&-(1-\Ad(u_\bi^\inv))\left(\Ad(v_0^\inv)\bx_0+\sum_{k=1}^{i-1}(1-\Ad(u_{A_k}))\bfh_{A_k}+(1-\Ad(u_{B_k}))\bfh_{B_k}\right)\nonumber\\
&+\Ad(u_\bi^\inv)\left((1-\Ad(u_\ai))\bfh_\ai+(1-\Ad(u_\bi))\bfh_\bi-\Ad(u_\bi)\bfh_\ai\right)\nonumber\\
\nonumber\\
\label{goodxs} \bfh_\ai=&\Ad(u_\ai^\inv u_\bi^\inv
u_\ai
u_{H_{i-1}}\cdots u_{H_1}v_0^\inv)\bx_\ai=\Ad(v_0^\inv v^\inv_{H_1}\cdots v^\inv_{H_{i-1}}v^\inv_\ai)\bx_\ai\\
\bfh_\bi=&-\Ad(u_\ai^\inv u_\bi^\inv u_\ai u_{H_{i-1}}\cdots
u_{H_1}v_0^\inv)\bx_\bi =-\Ad(v_0^\inv v^\inv_{H_1}\cdots
v^\inv_{H_{i-1}}v^\inv_\ai)\bx_\bi.\nonumber
\end{align}
Note that the variables  $\bfh_\ai,\bfh_\bi,\bl_\ai,\bl_\bi$ have
a clear geometrical interpretation. From Fig.~\ref{poly1} and
equation \eqref{cornergf} we see that $\bfh_\ai,\bfh_\bi$ can be
viewed as the parallel transport of $\bx_\ai,\bx_\bi$ from $p_0$
to the point $p_{4i-1}$, which is the starting point of the sides
$a'_i$ and $b'_i$ in the polygon $P_g$. Equivalently, we can
interpret them as the parallel transport of
$\Ad(v_\ai^\inv)\bx_\ai$ to $p_{4i-4}$ and of
$\Ad(v_\bi^\inv)\bx_\bi$ to $p_{4i-2}$, the starting points of
sides $a_i$ and $b_i$. Similarly, the variables $\bl_\ai$
represent the parallel transport of $\bj_\ai$ from  the starting
point $p_{4i-4}$ of side $a_i$, or, equivalently, of
$\Ad(u_\ai)\bj_\ai$ from its endpoint $p_{4i-3}$  to $p_0$, while
$\bl_\bi$ corresponds to the parallel transport  of $\bj_\bi$ from
$p_{4i-2}$ to $p_0$ or of $\Ad(u_\bi)\bj_\bi$ from $p_{4i-3}$ to
$p_0$.

Using expressions \eqref{holexplor} to \eqref{goodxs}, we can now
express the symplectic potential \eqref{gravtheta} on
$(\Poi)^{2g}$ in various combinations of the Lorentz and
translational components of holonomies and dual holonomies.
\newpage
\begin{theorem} $\quad$
\label{sympstructdualth}

\begin{enumerate}
\item In terms of the variables introduced in \eqref{holexplor} to
\eqref{goodxs}, the symplectic potential \eqref{gravtheta} is
given by
\begin{align} \label{gravtheta3}\Theta=&\sum_{i=1}^g\langle \bl_\ai ,v_\ai^\inv\delta
v_\ai\rangle+\langle\bl_\bi, v_\bi^\inv\delta
v_\bi\rangle\\=&\sum_{i=1}^g \langle \bfh_\ai, u_\ai^\inv\delta
u_\ai \rangle +\langle\bfh_\bi, u_\bi^\inv\delta
u_\bi\rangle+\langle \bj_\infty+\Ad(v_0^\inv)\bx_0,
u_\infty^\inv\delta u_\infty\rangle.\label{gravtheta6}
\end{align}

\item After a gauge transformation which acts on the holonomies
$N_\ai,N_\bi$ by simultaneous conjugation with the Poincar\'e
element  $(1, -\Ad(v_0)\bj_\infty-\bx_0)$
\begin{align}
&N_Y\mapsto \tilde{N}_Y\!=\!( v_Y, \tilde\bx_Y)
=(v_Y,\;\bx_Y\!\!-\!\!(1\!\!-\!\!\Ad(v_Y))(\Ad(v_0)\bj_\infty\!\!+\!\!\bx_0)\;
)\qquad Y\in\{A_1,...,B_g\}\nonumber\\
&\bfh_\ai\mapsto \tilde{\bfh}_\ai\!=\!\Ad( u_\ai^\inv u_\bi^\inv
u_\ai u_{H_{i-1}}\!\!\!\!\!\!\cdots \!u_{H_{1}} v_0^\inv)\tilde
\bx_\ai\nonumber\\
&\bfh_\bi\mapsto\tilde\bfh_\bi\!=\!-\!\Ad( u_\ai^\inv u_\bi^\inv
u_\ai u_{H_{i-1}}\!\!\!\!\!\!\cdots\! u_{H_{1}} v_0^\inv)\tilde
\bx_\bi,\label{tildbfhdef}
\end{align}
the symplectic potential \eqref{gravtheta} takes the form
\begin{align}
\label{gravtheta4}\Theta=&\sum_{i=1}^g\langle \tilde\bfh_\ai,
u_\ai^\inv\delta u_\ai\rangle+\langle \tilde\bfh_\bi,
u_\bi^\inv\delta u_\bi\rangle\\
\label{gravtheta5}=&\sum_{i=1}^g \langle
\Ad(v_\ai^\inv)\tilde\bx_\ai, \delta( v_{H_{i-1}}\cdots v_{H_1})(
v_{H_{i-1}}\cdots
v_{H_1})^\inv\rangle\\
& \qquad- \langle \Ad(v_\ai^\inv)\tilde\bx_\ai ,\delta(v_\ai^\inv
v_\bi^\inv v_\ai v_{H_{i-1}}\cdots v_{H_1})(v_\ai^\inv v_\bi^\inv
v_\ai v_{H_{i-1}}\cdots
v_{H_1})^\inv\rangle\nonumber\\
+&\sum_{i=1}^g\langle
\Ad(v_\bi^\inv)\tilde\bx_\bi,\delta(v_\ai^\inv v_\bi^\inv v_\ai
v_{H_{i-1}}\cdots v_{H_1})(v_\ai^\inv v_\bi^\inv v_\ai
v_{H_{i-1}}\cdots v_{H_1})^\inv \rangle\nonumber\\
&\qquad -\langle \Ad(v_\bi^\inv)\tilde\bx_\bi,\delta( v_\bi^\inv
v_\ai v_{H_{i-1}}\cdots v_{H_1})(v_\bi^\inv v_\ai
v_{H_{i-1}}\cdots v_{H_1})^\inv\rangle\nonumber.
\end{align}
\end{enumerate}
\end{theorem}
{\bf Proof:}

1. The proof is a straightforward but rather lengthy computation.
To prove \eqref{gravtheta3} we express the products of the Lorentz
components $u_\ai,u_\bi$ in \eqref{gravtheta} as products of
$v_\ai,v_\bi$
\begin{align}
\label{helpid1} &u_{H_{i-1}}\cdots u_{H_1}=v_0^\inv
v_{H_1}^\inv\cdots
v_{H_{i-1}}^\inv v_0\\
&u_\ai^\inv u_\bi^\inv u_\ai u_{H_{i-1}}\cdots u_{H_1}=v_0^\inv
v_{H_1}^\inv\cdots
v_{H_{i-1}}^\inv v_\ai^\inv v_0\nonumber\\
&u_\bi^\inv u_\ai u_{H_{i-1}}\cdots u_{H_1}=v_0^\inv
v_{H_1}^\inv\cdots v_{H_{i-1}}^\inv v_\ai^\inv v_\bi v_0\nonumber
\end{align}
and simplify the resulting products via the identity \bea
\label{derivid} \delta(a b)(a b)^\inv=\delta a a^\inv+\Ad(a)\delta
b b^\inv.  \eea Taking into account that the embedding of the
basepoint is not varied $\delta v_0=0$ and using the
$Ad$-invariance of the pairing $\langle\,,\,\rangle$ together with
\eqref{goodjs} we then obtain \eqref{gravtheta3}.

To prove \eqref{gravtheta6}, we insert expression \eqref{jxpr} for
the variables $\bj_\ai,\bj_\bi$ in terms of $\bfh_\ai,\bfh_\bi$
into \eqref{gravtheta} and isolate the terms containing
$\bfh_\ai,\bfh_\bi$. We then express the components of the
constraint \eqref{holconst} in terms of Lorentz and translational
components of the holonomies $\ai,\bi$ and $N_\ai, N_\bi$
according to
\begin{align}
\label{uinfexpr} u_\infty=&u_{H_g}\cdots u_{H_1}=v_0^\inv
v_{H_1}^\inv\cdots v_{H_g}^\inv v_0\\ \label{jinfexpr}
\bj_\infty=&\Ad(v_0^\inv) \sum_{i=1}^g
(1-\Ad(v_\ai))\bl_\ai+(1-\Ad(v_\bi))\bl_\bi\\
=&-(1-\Ad(u_\infty^\inv))\Ad(v_0^\inv)\bx_0+\Ad(u_\infty^\inv)\sum_{i=1}^g
(1-\Ad(u_\ai))\bfh_\ai+(1-\Ad(u_\bi))\bfh_\bi \nonumber.
\end{align}
Making use repeatedly of the identity \eqref{derivid} and of the
second identity in \eqref{jinfexpr} we  obtain \eqref{gravtheta3}.

2. Equation \eqref{gravtheta6} can be transformed into
\eqref{gravtheta4}, \eqref{gravtheta5} as follows. We first derive
an expression for the term $u_\infty^\inv\delta u_\infty$ in terms
of the Lorentz components $u_\ai,u_\bi$ from \eqref{uinfexpr}
\begin{align}
\label{uinfhelpid} &u_\infty^\inv\delta u_\infty=\\
&\sum_{i=1}^g \!\Ad(u_{H_1}^\inv\cdots
u_{H_{i-1}}^\inv)\!\left((1\!-\!\Ad(u_\ai^\inv u_\bi
u_\ai))u_\ai^\inv\delta u_\ai\!+\!(\Ad(u_\ai^\inv u_\bi
u_\ai)\!-\!\Ad(u_\ai^\inv u_\bi))u_\bi^\inv\delta
u_\bi\right).\nonumber
\end{align}
Expressing  $\bfh_\ai,\bfh_\bi$ in \eqref{gravtheta3} in terms of
$\tilde\bfh_\ai,\tilde\bfh_\bi$ and isolating the terms containing
$\bj_\infty+\Ad(v_0^\inv)\bx_0$ yields \eqref{gravtheta4}.
Finally, we express the Lorentz components $u_\ai,u_\bi$ in
\eqref{gravtheta4} as products in $v_\ai,v_\bi$ via
\eqref{holexplor}. After applying \eqref{tildbfhdef} and again
making use of \eqref{derivid} we obtain \eqref{gravtheta5}.
\hfill$\Box$

Thus, we find that the symplectic potential $\Theta$ takes a
particularly simple form when the components of the holonomies
$\ai,\bi$ are paired with those of $N_\ai,N_\bi$. Note also that
up to the term $\langle \bj_\infty+\Ad(v_0^\inv)\bx_0,
u_\infty^\inv\delta u_\infty\rangle$, which involves  the
components of the constraint \eqref{holconst} and can be
eliminated by performing the gauge transformation to the variables
$\tilde\bfh_\ai,\tilde\bfh_\bi$, the resulting expressions for the
symplectic potential are symmetric under the exchange
$\bl_\ai,\bl_\bi\leftrightarrow \bfh_\ai, \bfh_\bi$,
$v_\ai,v_\bi\leftrightarrow u_\ai,u_\bi$, which corresponds to
exchanging $\ai,\bi\leftrightarrow N_\ai,N_\bi$ and
$\gamma^\inv(p_0)\leftrightarrow \gamma(p_0)$. Similarly,
expression \eqref{gravtheta5} for the sympletic potential
 agrees with \eqref{gravtheta}, if we take into account the
difference in the parametrisation of the group elements $\ai, \bi$
and $N_\ai,N_\bi$ and exchange
$\bj_\ai\leftrightarrow\Ad(v_\ai)\tilde\bx_\ai$,
$\bj_\bi\leftrightarrow\Ad(v_\bi)\tilde\bx_\bi$. Hence, up to the
gauge transformation \eqref{tildbfhdef}, the symplectic potential
$\Theta$ takes the same form when expressed in terms of the
holonomies $\ai,\bi$ and in terms of $N_\ai,N_\bi$, as could be
anticipated from the symmetry in expressions \eqref{holexp3},
\eqref{holexp4}.

It follows from formula \eqref{gravtheta3} for the symplectic
potential that the only nontrivial Poisson brackets of the
variables $\bl_\ai,\bl_\bi$ and  $v_\ai,v_\bi$ are given by
\begin{align}
\label{pbls} &\{l^X_a,l^X_b\}=-\epsilon_{abc}l_X^c\quad  \{l^X_a,
v_X \}=- v_X J_a  & &X\in\{A_1,\ldots,B_g\}.
\end{align}
We can therefore identify the variables $l^X_a$ with the
left-invariant vector fields  defined as in \eqref{lorvecfs} and
acting on the copy of $\Lor$ associated to $v_X$
\begin{align}
\label{lvarvecs} &\{l_X^a,F\}(v_{A_1},...,
v_{B_g})\!\!=\!-J^a_{R_X}F(v_{A_1},...,
v_{B_g})\!\!=\!-\frac{d}{dt}F(v_{A_1},...,
v_Xe^{tJ_a}\!\!\!\!\!\!,...,v_{B_g})
\end{align}
for $F\in\cif((\Lor)^{2g})$, $X\in\{A_1,\ldots, B_g\}$. The same
holds for the Poisson brackets of $\tilde\bfh_\ai,\tilde\bfh_\bi$
with $u_\ai,u_\bi$
\begin{align}
\label{fvarvecs} &\{\tilde f_X^a,F\}(u_{A_1},...,
u_{B_g})\!\!=\!-J^a_{R_X}F(u_{A_1},...,
u_{B_g})\!\!=\!-\frac{d}{dt}F(u_{A_1},...,
u_Xe^{tJ_a}\!\!\!\!\!\!,...,u_{B_g}).
\end{align}

\subsection{Hamiltonians for grafting}
\label{hamsect}

We can now use the results from Sect.~\ref{dualpoiss} to show that
the mass $m_\lambda$ of a closed, simple curve
$\lambda\in\pi_1(S_g)$ generates the transformation of the
holonomies under grafting along $\lambda$.

\begin{theorem}
\label{masstheorem}

Consider a simple, closed curve
$\lambda=n_{x_r}^{\alpha_r}\circ\ldots\circ
n_{x_1}^{\alpha_1}\in\pi_1(S_g)$ and a general closed curve $\eta=
y_s^{\beta_s}\circ\ldots\circ y_1^{\beta_1}\in\pi_1(S_g)$ with
holonomies $H_\lambda$ and $H_\eta$, parametrised in terms of
$\ai,\bi$ and $N_{A_i}, N_{B_i}$ as
\begin{align}\label{twoholpar}
&H_\lambda=(u_\lambda,-\Ad(u_\lambda)\bj_\lambda)=\gamma(p_0)N_{X_r}^{\alpha_r}\cdots
N_{X_1}^{\alpha_1}\gamma(p_0)^\inv\\
&H_\eta=(u_\eta,-\Ad(u_\eta)\bj_\eta)=Y_s^{\beta_s}\cdots
Y_1^{\beta_1},\nonumber
\end{align}
where $X_i, Y_j\in\{A_1,\ldots,B_g\}$ and
$\alpha_i,\beta_j\in\{\pm 1\}$. Then, the transformation
\eqref{gencurvegraft} of the holonomy $H_\eta$ under grafting
along $\lambda$ is generated by the mass $m_\lambda$
\begin{align}
\label{poissgraft} &\{wm_\lambda, F\}=-\frac{d}{dt}|_{t=0} F\circ
Gr_{tw{\lambda}}\qquad F\in\cif((\Poi)^{2g})
\end{align}
where $ Gr_{tw{\lambda}}:(\Poi)^{2g}\rightarrow (\Poi)^{2g}$ is
given by  \eqref{graftaction}, \eqref{gencurvegraft}.
\end{theorem}

{\bf Proof:}

To prove the theorem, we calculate the Poisson bracket of
$\bp_\lambda^2=-m_\lambda^2$ with $\bj_\ai, \bj_\bi$. From
expression \eqref{pbls} for the Poisson bracket we have
\begin{align}
&\{l_\ai^a,
u_\lambda\}=-\!\!\!\!\!\!\!\!\!\sum_{X_k=\ai,\alpha_k=1}
\!\!\!\!\!\!\!u_\lambda\cdot\Ad(v_{X_{k-1}}^{\alpha_{k-1}}\cdots
v_{X_1}^{\alpha_1} v_0)^{ba}
J_b+\!\!\!\!\!\!\!\!\!\sum_{X_k=\ai,\alpha_k=-1}
\!\!\!\!\!\!\!u_\lambda\cdot\Ad(v_{X_{k}}^{\alpha_{k}}\cdots
v_{X_1}^{\alpha_1} v_0)^{ba} J_b\\
&\{l_\bi^a,
u_\lambda\}=-\!\!\!\!\!\!\!\!\!\sum_{X_k=\bi,\alpha_k=1}
\!\!\!\!\!\!\!u_\lambda\cdot\Ad(v_{X_{k-1}}^{\alpha_{k-1}}\cdots
v_{X_1}^{\alpha_1} v_0)^{ba}
J_b+\!\!\!\!\!\!\!\!\!\sum_{X_k=\bi,\alpha_k=-1}
\!\!\!\!\!\!\!u_\lambda\cdot\Ad(v_{X_{k}}^{\alpha_{k}}\cdots
v_{X_1}^{\alpha_1} v_0)^{ba} J_b.
\end{align}
Applying the formula \eqref{lorvecfs} for the left-invariant
vector fields on $\Lor$ to $F=\bp_\lambda^2$ yields
\begin{align}
\label{lgrbrackets} &\{\bl_\ai,
\bp_\lambda^2\}=2\!\!\!\!\!\!\sum_{X_k=\ai,\alpha_k=1}\!\!\!\!\!\!\Ad(v_{X_{k-1}}^{\alpha_{k-1}}\cdots
v^{\alpha_1}_{X_1}v_0)\bp_\lambda-2\!\!\!\!\!\!\sum_{X_k=\ai,\alpha_k=-1}\!\!\!\!\!\!\Ad(v^{\alpha_k}_{X_{k}}\cdots v^{\alpha_1}_{X_1}v_0)\bp_\lambda\\
&\{\bl_\bi,
\bp_\lambda^2\}=2\!\!\!\!\!\!\sum_{X_k=\bi,\alpha_k=1}\!\!\!\!\!\!\Ad(v_{X_{k-1}}^{\alpha_{k-1}}\cdots
v_{X_1}^{\alpha_1}v_0)\bp_\lambda-2\!\!\!\!\!\!\sum_{X_k=\bi,\alpha_k=-1}\!\!\!\!\!\!\Ad(v_{X_{k}}^{\alpha_k}\cdots
v_{X_1}^{\alpha_1}v_0)\bp_\lambda\nonumber,
\end{align}
where the expressions involving vectors are to be understood
componentwise. With expression \eqref{goodjs} relating $\bl_\ai,
\bl_\bi$ to $\bj_\ai,\bj_\bi$ and setting
$\hat\bp_\lambda=\frac{1}{m_\lambda}\bp_\lambda$,
$\bp_\lambda^2=-m_\lambda^2$, we obtain
\begin{align}
\label{jgrbrackets2} &\{ m_\lambda, \bj_\ai\}\!\!=\!\Ad(v_0^\inv
v_{H_1}^\inv\cdots v^\inv_{H_{i-1}}) \!\left(\!\!\!\!
\!\!\!\!\!\!\!\!\!\!\!\!\sum_{\quad\qquad
X_k=\ai,\alpha_k=1}\!\!\!\!\!\!\!\!\!\!\!\!\!\!\!\!\!\Ad(
v_{X_{k-1}}^{\alpha_{k-1}}\!\!\!\!\!\!\cdots
v_{X_1}^{\alpha_{1}}v_0)\hat \bp_\lambda
\!-\!\!\!\!\!\!\!\!\!\!\!\!\sum_{X_k=\ai,\alpha_k=-1}\!\!\!\!\!\!\!\!\!\!\!\Ad(
v_{X_{k}}^{\alpha_{k}}\!\!\!\!\!\!\cdots
v_{X_1}^{\alpha_{1}}v_0)\hat \bp_\lambda \right)
\\
&\{m_\lambda, \bj_\bi\}\!\!=\!-\Ad(v_0^\inv v_{H_1}^\inv\cdots
v^\inv_{H_{i-1}}v_\ai^\inv v_\bi)\!\left(\!\!\!\!
\!\!\!\!\!\!\!\!\!\!\!\!\sum_{\quad\qquad
X_k=\bi,\alpha_k=1}\!\!\!\!\!\!\!\!\!\!\!\!\!\!\!\!\Ad(
v_{X_{k-1}}^{\alpha_{k-1}}\!\!\!\!\!\!\cdots
v_{X_1}^{\alpha_{1}}v_0)\hat \bp_\lambda
\!-\!\!\!\!\!\!\!\!\!\!\!\!\sum_{X_k=\bi,\alpha_k=-1}\!\!\!\!\!\!\!\!\!\!\Ad(
v_{X_{k}}^{\alpha_{k}}\!\!\!\!\!\!\cdots
v_{X_1}^{\alpha_{1}}v_0)\hat \bp_\lambda \right)\nonumber.
\end{align}
Using expression \eqref{etavect} for the variable
$\bj_\eta$ as a linear combination of $\bj_\ai,\bj_\bi$ and taking
into account
 the relation \eqref{nprel} between the vector $\hat\bp_\lambda$ and the vector $\bn_{p,q}$ in \eqref{gencurvegraft}, we find agreement
with \eqref{gencurvegraft} up to a sign, which proves
\eqref{poissgraft} . \hfill$\Box$

Hence, we find that the transformation of the holonomies under
grafting along a closed, simple geodesic $\lambda$ on $S_\Gamma$
is generated by the mass $m_\lambda$. Note, however, that the
transformation generated by the mass $m_\lambda$
 is defined for general closed curves $\lambda\in\pi_1(S_g)$ and as
a map $(\Poi)^{2g}\rightarrow (\Poi)^{2g}$. In contrast, the
grafting procedure defined in \cite{bg,bb} whose action on the
holonomies is given in Sect.~\ref{Csgraft} is defined for simple,
closed curves and acts on static spacetimes for which the
translational components of the dual holonomies $N_\ai,N_\bi$
vanish and their Lorentz components are the generators of a
cocompact Fuchsian group $\Gamma$. In this sense, the
transformation $Gr_{\lambda}:(\Poi)^{2g}\rightarrow (\Poi)^{2g}$
generated by the mass $m_\lambda$ can be viewed as an extension of
the grafting procedure in \cite{bg,bb} to non-simple curves and to
the whole Poisson manifold $((\Poi)^{2g},\Theta)$.

The fact that the transformation of the holonomies under grafting
 is generated
via the Poisson bracket allows us to deduce some properties of
this transformation which would be much less apparent from the
general formula \eqref{gencurvegraft}.

\begin{corollary} $\quad$
\label{graftprop}

\begin{enumerate}
\item The action of grafting leaves the constraint
\eqref{holconst} invariant and commutes with the associated gauge
transformation by simultaneous conjugation
\begin{align}
\label{constcomm} \{u_\infty, m_\lambda\}=\{j_\infty^a,
m_\lambda\}=0.
\end{align}

\item  The grafting transformations $Gr_{w_i\lambda_i}$ for
different closed curves $\lambda_i\in\pi_1(S_g)$ with weights
$w_i\in\RR^+$ commute and satisfy
\begin{align}
\label{graftcomm} \{\sum_{i=1}^n w_i m_{\lambda_i},
F\}=-\frac{d}{dt}|_{t=0}F\circ Gr_{tw_n\lambda_n}\circ\ldots\circ
Gr_{tw_1\lambda_1}\qquad F\in\cif((\Poi)^{2g}).
\end{align}

\item The grafting maps $Gr_{w\lambda}$ act on the Poisson
manifold $((\Poi)^{2g}, \Theta)$ via Poisson isomorphisms
\begin{align}
\label{poissisom} \{F\circ Gr_{w\lambda}, G\circ
Gr_{w\lambda}\}=\{F,G\}\circ Gr_{w\lambda}\qquad
F,G\in\cif((\Poi)^{2g}).
\end{align}
\end{enumerate}
\end{corollary}

{\bf Proof:}  That the components of the constraint
\eqref{holconst} Poisson commute with the mass $m_\lambda$ follows
from the fact that $m_\lambda$ is an observable of the theory, but
can also be checked by direct calculation. It is shown in
\cite{we1} that the components $j_\infty^a$ act on the Lorentz
components $u_\ai, u_\bi$ by simultaneous conjugation with $\Lor$,
which leaves all masses $m_\lambda$ invariant.

To prove the second statement, we recall that all Lorentz
components $u_\ai,u_\bi$ Poisson commute, which together with
\eqref{gencurvegraft} and \eqref{poissgraft} implies the
commutativity of grafting. Differentiating then yields
\eqref{graftcomm}.

The third statement follows  directly from the fact that the
grafting transformation is generated via the Poisson bracket by a
standard argument making use of the Jacobi identity. In our case,
the fact that the Lorentz components $u_\ai,u_\bi$ Poisson commute
allows one to write
\begin{align}
\{F\circ Gr_{w\lambda}, G\circ
Gr_{w\lambda}\}=&\!\!\!\!\!\!\sum_{\!\!\!\!\!\!\!\!\!\!\!\!\!\!\!\!\!\!\!\!\!\!\!\!X,Y\in\{A_1,\ldots,B_g\}}\frac{\partial
 F}{\partial j_X^a} \frac{\partial G}{\partial j^b_Y}\left(\{j_X^a, j_Y^b\}-w\{\{m_\lambda, j_X^a\},
 j_Y^b\}-w\{j_X^a,
 \{m_\lambda,j_Y^b\}\}\right)\nonumber,
\end{align}
and, using the Jacobi identity for the last two brackets, one
obtains \eqref{poissisom}
\begin{align}
\{F\circ Gr_{w\lambda}, G\circ
Gr_{w\lambda}\}=&\!\!\!\!\!\!\sum_{\!\!\!\!\!\!\!\!\!\!\!\!\!\!\!\!\!\!\!\!\!\!\!\!X,Y\in\{A_1,\ldots,B_g\}}\frac{\partial
 F}{\partial j_X^a} \frac{\partial G}{\partial j^b_Y}\left(\{j_X^a, j_Y^b\}\!\!-\!\!w\{m_\lambda, \{j_X^a,
 j_Y^b\}\}\right)=\{F,G\}\circ Gr_{w\lambda}\;\;\;\;\Box\nonumber
\end{align}


After deriving the Hamiltonians that generate the transformation
of the holonomies under grafting along a closed, simple curve
$\lambda\in\pi_1(S_g)$, we will now demonstrate that Theorem
\ref{masstheorem} gives rise to a  general symmetry relation
between the Poisson brackets of mass and spin associated to
general closed curves $\lambda,\eta\in\pi_1(S_g)$.
\begin{theorem}
\label{dualtheorem} The Poisson brackets of mass and spin for
$\lambda,\eta\in \pi_1(S_g)$ satisfy the relation
\begin{align}
\label{dualrel} &\{\bp_\eta\cdot\bj_\eta,
\bp^2_\lambda\}=\{\bp^2_\eta,\bp_\lambda\cdot\bj_\lambda\} &
&\{m_\eta, s_\lambda\}=\{s_\eta, m_\lambda\}.
\end{align}
\end{theorem}

{\bf Proof:} To prove \eqref{dualrel}, we consider curves
$\lambda,\eta\in\pi_1(S_g)$  with holonomies $H_\lambda, H_\eta$
parametrised as in \eqref{twoholpar}. From \eqref{lgrbrackets}
 it follows that the Poisson bracket of $\bp_\eta\cdot \bj_\eta$
and $\bp^2_\lambda$ is given by
\begin{align}
\label{masspinbr1} &\{\bp_\eta\cdot\bj_\eta,
\bp_\lambda^2\}=2\sum_{i=1}^g\left(\!\!\!\!\!\!\!\!\!\!\!\sum_{\qquad
Y_k=\ai,\beta_k=1}\!\!\!\!\!\!\!\!\!\!\!\Ad(u^{\beta_{k-1}}
_{Y_{k-1}}\cdots
u_{Y_1}^{\beta_1})\bp_\eta-\!\!\!\!\!\!\!\!\!\!\!\sum_{Y_k=\ai,\beta_k=-1}\!\!\!\!\!\!\!\!\!\!\!\Ad(u^{\beta_{k}}_{Y_{k}}\cdots
u_{Y_1}^{\beta_1})\bp_\eta\right)\cdot\\
&\left(\!\!\!\!\!\!\!\!\!\!\!\sum_{\qquad
X_k=\ai,\alpha_k=1}\!\!\!\!\!\!\!\!\!\!\!\Ad(v_0^\inv
v^\inv_{H_1}\cdots v^\inv_{H_{i-1}}
v_{X_{k-1}}^{\alpha_{k-1}}\cdots
v_{X_1}^{\alpha_{1}}v_0)\bp_\lambda-\!\!\!\!\!\!\!\!\!\!\!\sum_{X_k=\ai,\alpha_k=-1}\!\!\!\!\!\!\!\!\!\!\!\Ad(v_0^\inv
v^\inv_{H_1}\cdots v^\inv_{H_{i-1}} v_{X_{k}}^{\alpha_{k}}\cdots
v_{X_1}^{\alpha_{1}}v_0)\bp_\lambda\right)\nonumber\\
-&2\sum_{i=1}^g\left(\!\!\!\!\!\!\!\!\!\!\!\sum_{\qquad
Y_k=\bi,\beta_k=1}\!\!\!\!\!\!\!\!\!\!\!\Ad(u^{\beta_{k-1}}
_{Y_{k-1}}\cdots
u_{Y_1}^{\beta_1})\bp_\eta-\!\!\!\!\!\!\!\!\!\!\!\sum_{Y_k=\bi,\beta_k=-1}\!\!\!\!\!\!\!\!\!\!\!\Ad(u^{\beta_{k}}_{Y_{k}}\cdots
u_{Y_1}^{\beta_1})\bp_\eta\right)\cdot\nonumber\\
&\left(\!\!\!\!\!\!\!\!\!\!\!\!\!\!\!\!\!\!\sum_{\quad\qquad
X_k=\bi,\alpha_k=1}\!\!\!\!\!\!\!\!\!\!\!\!\!\!\!\!\!\!\!\!\!\!\!\Ad(v_0^\inv
v^\inv_{H_1}\cdots v^\inv_{H_{i-1}}v_\ai^\inv v_\bi
v_{X_{k-1}}^{\alpha_{k-1}}\cdots
v_{X_1}^{\alpha_{1}}v_0)\bp_\lambda\!\!\!-\!\!\!\!\!\!\!\!\!\!\!\!\!\!\!\!\!\!\!\!\!\!\sum_{\quad\qquad
X_k=\bi,\alpha_k=-1}\!\!\!\!\!\!\!\!\!\!\!\!\!\!\!\!\!\!\!\!\!\!\!\!\!\Ad(v_0^\inv
v^\inv_{H_1}\cdots v^\inv_{H_{i-1}}v_\ai^\inv v_\bi
v_{X_{k}}^{\alpha_{k}}\cdots
v_{X_1}^{\alpha_{1}}v_0)\bp_\lambda\right).\nonumber
\end{align}
To compute the Poisson bracket
$\{\bp^2_\eta,\bp_\lambda\cdot\bj_\lambda\}$, we express the
translational component of the  holonomy $H_\lambda$ in terms of
the holonomies $N_\ai,N_\bi$
\begin{align}
\label{jlamdexpr} \bj_\lambda=\!-\!\!\!\!\sum_{k:
\alpha_k=1}\!\!\Ad(v_0^\inv v_{X_1}^{-\alpha_1}\cdots
v_{X_{k-1}}^{-\alpha_{k-1}}v_{X_k}^\inv)\bx_{X_k}+\!\!\!\!\sum_{k:
\alpha_k=-1}\!\!\Ad(v_0^\inv v_{X_1}^{-\alpha_1}\cdots
v_{X_{k}}^{-\alpha_{k}}v_{X_k}^\inv)\bx_{X_k}.
\end{align}As simultaneous
conjugation of all holonomies with a general Poincar\'e valued
function on $(\Poi)^{2g}$ leaves $\bp_\lambda\cdot\bj_\lambda$
invariant, we can replace $\bx_\ai,\bx_\bi$ by
$\tilde\bx_\ai,\tilde \bx_\bi$ in expression \eqref{jlamdexpr}.
Using expression \eqref{fvarvecs} for the Poisson bracket of
$\tilde\bfh_\ai,\tilde\bfh_\bi$ with $u_\ai,u_\bi$, equation
\eqref{tildbfhdef} relating $\tilde\bfh_\ai,\tilde\bfh_\bi$ and
$\tilde\bx_\ai,\tilde\bx_\bi$ and expression \eqref{lorvecfs} for
the action of the left-invariant vector fields on $\Lor$, we find
that the Poisson bracket of $\tilde\bx_\ai,\tilde\bx_\bi$ with
$\bp_\lambda^2$ is given by
\begin{align}
&\{\tilde \bx_\ai, \bp^2_\eta\}=2\Ad( v_\ai v_{H_{i-1}}\cdots
v_{H_1} v_0) \left(\!\!\!\!\!\!\!\!\!\!\!\sum_{\qquad
Y_k=\ai,\beta_k=1}\!\!\!\!\!\!\!\!\!\!\!\Ad(u^{\beta_{k-1}}
_{Y_{k-1}}\cdots
u_{Y_1}^{\beta_1})\bp_\eta-\!\!\!\!\!\!\!\!\!\!\!\sum_{Y_k=\ai,\beta_k=-1}\!\!\!\!\!\!\!\!\!\!\!\Ad(u^{\beta_{k}}_{Y_{k}}\cdots
u_{Y_1}^{\beta_1})\bp_\eta\right)\nonumber\\
&\{\tilde \bx_\bi, \bp^2_\eta\}=-2\Ad(v_\ai v_{H_{i-1}}\cdots
v_{H_1} v_0 )\left(\!\!\!\!\!\!\!\!\!\!\!\sum_{\quad
Y_k=\bi,\beta_k=1}\!\!\!\!\!\!\!\!\!\!\!\Ad(u^{\beta_{k-1}}
_{Y_{k-1}}\cdots
u_{Y_1}^{\beta_1})\bp_\eta-\!\!\!\!\!\!\!\!\!\!\!\sum_{Y_k=\bi,\beta_k=-1}\!\!\!\!\!\!\!\!\!\!\!\Ad(u^{\beta_{k}}_{Y_{k}}\cdots
u_{Y_1}^{\beta_1})\bp_\eta\right).\nonumber
\end{align}
Replacing $\bx_\ai\rightarrow \tilde\bx_\ai$,
$\bx_\bi\rightarrow\tilde\bx_\bi$ in expression \eqref{jlamdexpr}
for $\bj_\eta$ then yields
\begin{align}
\label{masspinbr2} &\{\bp_\eta^2,
\bj_\lambda\}=2\sum_{i=1}^g\left(\!\!\!\!\!\!\!\!\!\!\!\sum_{\qquad
X_k=\ai,\alpha_k=1}\!\!\!\!\!\!\!\!\!\!\!\Ad(v_0^\inv
v_{X_1}^{-\alpha_{1}}\cdots
v_{X_{k-1}}^{-\alpha_{k-1}})-\!\!\!\!\!\!\!\!\!\!\!\sum_{X_k=\ai,\alpha_k=-1}\!\!\!\!\!\!\!\!\!\!\!\Ad(v_0^\inv
v_{X_1}^{-\alpha_{1}}\cdots
v_{X_{k}}^{-\alpha_{k}})\right)\\
&\Ad(v_{H_{i-1}}\cdots
v_{H_1}v_0)\left(\!\!\!\!\!\!\!\!\!\!\!\sum_{\qquad
Y_k=\ai,\beta_k=1}\!\!\!\!\!\!\!\!\!\!\!\Ad(u^{\beta_{k-1}}
_{Y_{k-1}}\cdots
u_{Y_1}^{\beta_1})\bp_\eta-\!\!\!\!\!\!\!\!\!\!\!\sum_{Y_k=\ai,\beta_k=-1}\!\!\!\!\!\!\!\!\!\!\!\Ad(u^{\beta_{k}}_{Y_{k}}\cdots
u_{Y_1}^{\beta_1})\bp_\eta\right)\nonumber\\
-&2\sum_{i=1}^g\left(\!\!\!\!\!\!\!\!\!\!\!\sum_{\qquad
X_k=\bi,\alpha_k=1}\!\!\!\!\!\!\!\!\!\!\!\Ad(v_0^\inv
v_{X_1}^{-\alpha_{1}}\cdots
v_{X_{k-1}}^{-\alpha_{k-1}})-\!\!\!\!\!\!\!\!\!\!\!\sum_{X_k=\bi,\alpha_k=-1}\!\!\!\!\!\!\!\!\!\!\!\Ad(v_0^\inv
v_{X_1}^{-\alpha_{1}}\cdots
v_{X_{k}}^{-\alpha_{k}})\right)\nonumber\\
&\Ad(v_\bi^\inv v_\ai v_{H_{i-1}}\cdots
v_{H_1}v_0)\left(\!\!\!\!\!\!\!\!\!\!\!\sum_{\qquad
Y_k=\bi,\beta_k=1}\!\!\!\!\!\!\!\!\!\!\!\Ad(u^{\beta_{k-1}}
_{Y_{k-1}}\cdots
u_{Y_1}^{\beta_1})\bp_\eta-\!\!\!\!\!\!\!\!\!\!\!\sum_{Y_k=\bi,\beta_k=-1}\!\!\!\!\!\!\!\!\!\!\!\Ad(u^{\beta_{k}}_{Y_{k}}\cdots
u_{Y_1}^{\beta_1})\bp_\eta\right),\nonumber
\end{align}
and multiplication with $\bp_\lambda$ gives
\eqref{masspinbr1}.\hfill $\Box$

The geometrical implications of Theorem \ref{dualtheorem} are that
the change of the spin $s_\eta$ of a closed, simple curve
$\eta\in\pi_1(S_g)$ under grafting along another closed, simple
curve $\lambda\in\pi_1(S_g)$ is the same as the change of the spin
$s_\lambda$ under grafting along $\eta$. Furthermore, it is shown
in \cite{we3}, for a summary of the results see
Sect.~\ref{grdtrel}, that the product $m_\lambda s_\lambda$ of
mass and spin of a closed, simple curve $\lambda\in\pi_1(S_g)$ is
the Hamiltonian which generates an infinitesimal Dehn twist around
$\lambda$. Thus, Theorem \ref{dualtheorem} implies that
 the transformation of the
mass $m_\eta$ under an infinitesimal Dehn twist around
$\lambda\in\pi_1(S_g)$ agrees with the transformation of the spin
$s_\eta$ under infinitesimal grafting along $\lambda$. We will
clarify this connection further in the next section, where we
discuss the relation between grafting and Dehn twists.

\section{Grafting and Dehn twists}
\label{grdtrel}  In this section, we show that there is a link
between the transformation of the holonomies  under grafting and
under Dehn twists along a general closed, simple curve
$\lambda\in\pi_1(S_g)$.

The transformation of the holonomies under Dehn twists is
investigated in \cite{we3} for Chern-Simons theory on a manifold
of topology $\RR\times S_{g,n}$, where $S_{g,n}$ is a general
orientable two-surface of genus $g$ with $n$ punctures. The gauge
groups considered in \cite{we3} are of the form
$G\ltimes\mathfrak{g}^*$, where $G$ is a finite dimensional,
connected, simply connected and unimodular Lie group,
$\mathfrak{g}^*$ the dual of its Lie algebra and $G$ acts on
$\mathfrak{g}^*$ in the coadjoint representation. The assumption
of simply-connectedness in \cite{we3} gives rise to technical
simplifications in the quantised theory but does not affect the
classical results. Hence, reasoning and results in \cite{we3}
apply to the case of gauge group $\Poi$ and can be summarised as
 follows.

\begin{theorem}\cite{we3}
\label{dttheorem}

For any simple, closed curve $\lambda\in\pi_1(S_g)$ with holonomy
$H_\lambda=(u_\lambda, -\Ad(u_\lambda)\bj_\lambda)$,
$u_\lambda=e^{-p_\lambda^a J_a}$,  the product of the associated
mass and spin $\bp_\lambda\cdot\bj_\lambda=m_\lambda s_\lambda$
generates an infinitesimal Dehn twist around $\lambda$ via the
Poisson bracket defined by \eqref{gravtheta}
\begin{align}
\label{stpb} \{m_\lambda s_\lambda, F\}=\frac{d}{dt}|_{t=0} F\circ
D_{t\lambda}\qquad F\in\cif((\Poi)^{2g}),
\end{align}
where $D_{t\lambda}:(\Poi)^{2g}\rightarrow (\Poi)^{2g}$ agrees
with the action $D_\lambda:(\Poi)^{2g}\rightarrow (\Poi)^{2g}$ of
the Dehn-twist around $\lambda$ for $t=1.$ The transformation
$D_{t\lambda}$ acts on the Poisson manifold $((\Poi)^{2g},\Theta)$
via Poisson isomorphisms
\begin{align}
\label{dtpoiss} \{F\circ D_{t\lambda}, G\circ
D_{t\lambda}\}=\{F,G\}\circ D_{t\lambda}\qquad
F,G\in\cif((\Poi)^{2g}).
\end{align}
\end{theorem}
As in the definition of the grafting map
$Gr_{w\lambda}:(\Poi)^{2g}\rightarrow(\Poi)^{2g}$, the different
copies of $\Poi$ in Theorem \ref{dttheorem} stand for the
holonomies $\ai,\bi$. However, unlike our derivation of the
grafting map, the derivation in \cite{we3} does not make use of
the dual generators $n_{a_i}, n_{b_i}$ but is formulated entirely
in terms of the holonomies $\ai,\bi$. The action
$D_{t\lambda}:(\Poi)^{2g}\rightarrow (\Poi)^{2g}$ of
(infinitesimal) Dehn twists on the holonomies is determined
graphically. As this graphical procedure will play an important
role in relating Dehn twists and grafting, we present it here in a
slightly different and more
 detailed version than in \cite{we3}.

We consider simple curves $\lambda,\eta\in\pi_1(S_g)$ parametrised
in terms of the generators $a_i,b_i\in\pi_1(S_g)$ as $\lambda
=z_t^{\delta_t}\circ\ldots\circ z_1^{\delta_1}$,
$\eta=y_s^{\beta_s}\circ\ldots\circ y_1^{\beta_1}$ with $z_i,
y_j\in\{a_1,\ldots,b_g\}$, $\beta_j,\delta_k \in\{\pm1\}$ and
associated holonomies
\begin{align}
\label{singlehols} &H_\lambda=Z_t^{\delta_t}\cdots
Z_1^{\delta_1}=e^{-(p_\lambda^a+\theta
k_\lambda^a)J_a}=(u_\lambda,-\Ad(u_\lambda)\bj_\lambda)\\
&H_\eta=Y_s^{\beta_s}\cdots Y_1^{\beta_1} =e^{-(p_\eta^a+\theta
k_\eta^a)J_a}=(u_\eta,-\Ad(u_\eta)\bj_\eta)\nonumber.
\end{align}
 To determine the action of the transformation generated by
$m_\lambda s_\lambda$ on the holonomy $H_\eta$, we consider the
surface $S_g-D$ obtained from $S_g$ by removing a
disc\footnote{The reason for the removal of the disc is that we
work on an extended phase space where the constraint
\eqref{holconst} arising from the defining relation of the
fundamental group is not imposed. It is shown in \cite{we3} that
this implies that instead of the mapping class group
$\text{Map}(S_g)$, it is the  mapping class group
$\text{Map}(S_g-D)$ that acts on the Poisson manifold
$((\Poi)^{2g},\Theta)$.} $D$.

We represent the generators $a_i,b_i\in\pi_1(S_g)$ by curves as in
Fig.~\ref{pi1gr}, but instead of a basepoint, we draw a line on
which the starting points $s_{a_i}, s_{b_i}$ and endpoints
$t_{a_i}, t_{b_i}$ are ordered\footnote{This ordering corresponds
to an ordering of the edges at each vertex needed to define the
Poisson structure in the formalism developed by Fock and Rosly
\cite{FR}.} (from right to left) according to
\begin{align}
\label{pi1genord}
s_{a_1}<s_{b_1}<t_{a_1}<t_{b_1}<s_{a_2}<s_{b_2}<t_{a_2}<t_{b_2}<\ldots<s_{a_g}<s_{b_g}<t_{a_g}<t_{b_g}.
\end{align}
The curves representing the generators $a_i,b_i\in\pi_1(S_g)$
start and end in, respectively, $s_{a_i}, s_{b_i}$ and $t_{a_i},
t_{b_i}$ and their inverses in
$s_{a_i^\inv}=t_{a_i}$,$s_{b_i^\inv}=t_{b_i}$ and
$t_{a_i^\inv}=s_{a_i}$,$t_{b_i^\inv}=s_{b_i}$.

\begin{figure}
\vskip .3in \protect\input epsf \protect\epsfxsize=12truecm
\protect\centerline{\epsfbox{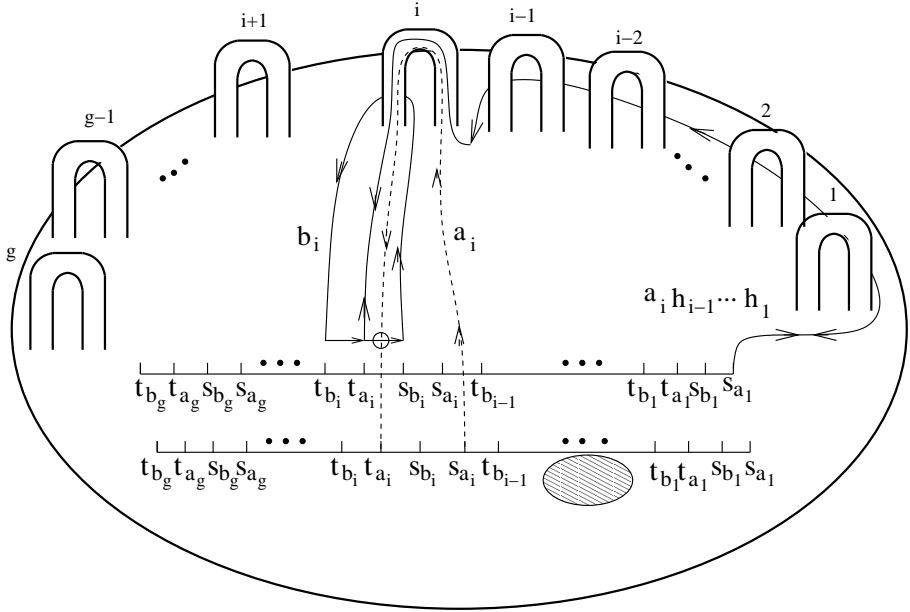}} \caption{The
decomposition of $n_{a_i}=h_1^\inv\circ\ldots\circ h_{i-1}^\inv
a_i^\inv \circ b_i\circ a_i\circ h_{i-1}\circ\ldots\circ h_1$
(full line) and its intersection with $a_i$ (dashed line);
segments in the decomposition of $n_{a_i}$ that do not intersect
any generator $a_j,b_j\in\pi_1(S_g)$ are omitted} \label{inters1}
\end{figure}

\begin{figure}
\vskip .3in \protect\input epsf \protect\epsfxsize=12truecm
\protect\centerline{\epsfbox{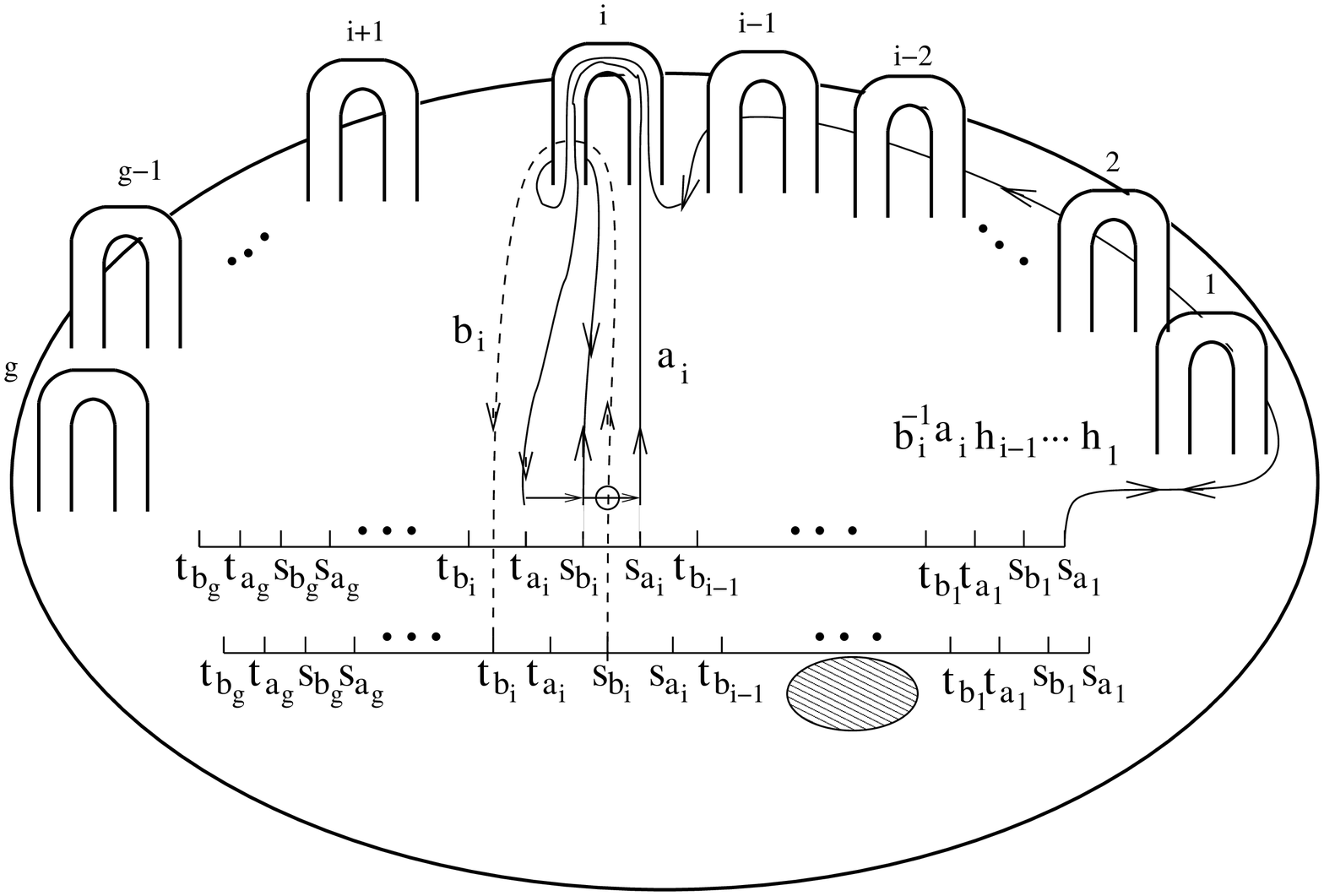}} \caption{The
decomposition of $n_{b_i}=h_1^\inv\circ\ldots\circ h_{i-1}^\inv
a_i^\inv \circ b_i\circ a_i \circ b_i^\inv\circ a_i\circ
h_{i-1}\circ\ldots\circ h_1$ (full line) and its intersection with
$b_i$ (dashed line); segments in the decomposition of
$n_{b_i}$that do not intersect any generator
$a_j,b_j\in\pi_1(S_g)$ are omitted} \label{inters2}
\end{figure}

To derive the transformation of the holonomy $H_\eta$ under an
(infinitesimal) Dehn twist along $\lambda$, we draw two such
lines, one corresponding to $\eta$, one to $\lambda$ such that the
line for $\eta$ is tangent to the disc, while the one for
$\lambda$ is displaced slightly away from the disc. We then
decompose the curves representing $\eta$ and $\lambda$ graphically
 into the curves representing the
generators $a_i,b_i$ and their inverses,  with ordered starting
and end points on the corresponding lines, and into segments
parallel to the lines which connect the starting and endpoints of
different factors, see Fig.~\ref{inters1}, Fig.~\ref{inters2},
Fig.~\ref{exdt}, Fig.~\ref{exdt2}.  The curves representing
$a_i^{\pm 1}, b_i^{\pm 1}$ and the segments connecting their
starting and endpoints are drawn in such a way that there is a
minimal number of intersection points and such that all
intersection points occur on the lines connecting different
starting and endpoints of generators $a_i,b_i$ in the
decomposition  of $\lambda$, as shown in Fig.~\ref{inters1},
Fig.~\ref{inters2}, Fig.~\ref{exdt}, Fig.~\ref{exdt2}. An
intersection point $q_i$ is said to occur between the factors
$z_i^{\delta_i}$ and $z_{i+1}^{\delta_{i+1}}$ on $\lambda$ if it
lies on the straight line connecting $t_{z_i^{\delta_i}}$ and
$s_{z_{i+1}^{\delta_{i+1}}}$, where
$z_{t+1}^{\delta_{t+1}}=z_1^{\delta_1}$. Similarly,  an
intersection point occurs between the factors $y_i^{\beta_i}$ and
$y_{i+1}^{\beta_{i+1}}$ on $\eta$ if it lies on
$y_{i+1}^{\beta_{i+1}}$ near the starting point
$s_{y_{i+1}^{\beta_{i+1}}}$ or on $y_{i}^{\beta_{i}}$ near the
endpoint  $t_{y_i^{\beta_i}}$.

Let now $\lambda,\eta\in\pi_1(S_g)$  have intersection points
$q_1,\ldots, q_n$ such that $q_i$ occurs between
$z_{k_i}^{\delta_{k_i}}$ and $z_{k_i+1}^{\delta_{k_i+1}}$ on
$\lambda$  and between $y_{j_i}^{\beta_{j_i}}$ and
$y_{j_i+1}^{\beta_{j_{i}+1}}$ on $\eta$  with $j_1\leq j_2\leq
\ldots\leq j_n$. We denote by
$\epsilon_i=\epsilon_i(\lambda,\eta)$ the oriented intersection
number in $q_i$ with the convention $\epsilon_i=1$ if $\lambda$
crosses $\eta$ from the left to the right in the direction of
$\eta$. It is shown in \cite{we3} that, with these conventions,
the action of an infinitesimal Dehn twist $D_{t\lambda}:
(\Poi)^{2g}\rightarrow(\Poi)^{2g}$ is given by inserting the
Poincar\'e element
\begin{align}
\label{insertel}  ((Z_{k_i}^{\delta_{k_i}}
Z_{k_{i}-1}^{\delta_{k_{i}-1}}\!\!\cdots\!
Z_{1}^{\delta_1})H_\lambda(Z_{k_i}^{\delta_{k_i}}
Z_{k_{i}-1}^{\delta_{k_{i}-1}}\!\!\cdots\!
Z_{1}^{\delta_1})^\inv)^{t\epsilon_i}\!\!=\!
(Z_{k_i}^{\delta_{k_i}} \!\!\cdots\! Z_{1}^{\delta_1})
e^{-t\epsilon_i (p^a_\lambda +\theta k^a_\lambda)J_a}\!
(Z_{k_i}^{\delta_{k_i}} \!\!\cdots\! Z_{1}^{\delta_1})^\inv
\end{align}
between the factors $Y_{j_i}^{\beta_{j_i}}$ and
$Y_{j_i+1}^{\beta_{j_{i}+1}}$
\begin{align}
\label{dtdef} D_{t\lambda}:\;H_\eta\mapsto &Y^{\beta_s}_s\cdots
Y^{\beta_{j_n+1}}_{j_n+1}\cdot (Z_{k_n}^{\delta_{k_n}}\cdots
Z_1^{\delta_1}) e^{-t\epsilon_n(p_\lambda^a+\theta
k_\lambda^a)J_a} (Z_{k_n}^{\delta_{k_n}}\cdots
Z_1^{\delta_1})^\inv\cdot Y^{\beta_{j_n}}_{j_n}\cdots
Y^{\beta_{j_{n-1}+1}}_{j_{n-1}+1}\cdot\nonumber\\
\cdot &(Z_{k_{n-1}}^{\delta_{k_{n-1}}}\!\!\cdots Z_1^{\delta_1})
e^{-t\epsilon_{n-1}(p_\lambda^a+\theta k_\lambda^a)J_a}
(Z_{k_{n-1}}^{\delta_{k_{n-1}}}\!\!\cdots Z_1^{\delta_1})^\inv
Y^{\beta_{j_{n-1}}}_{j_{n-1}}\!\!\cdots
Y^{\beta_{j_{n-2}+1}}_{j_{n-2}+1}()\cdots()\cdot\nonumber\\
\cdot &Y^{\beta_{j_{2}}}_{j_{2}}\!\!\cdots
Y^{\beta_{j_1+1}}_{j_1+1}\cdot(Z_{k_{1}}^{\delta_{k_1}}\cdots
Z_1^{\delta_1}) e^{-t\epsilon_1(p_\lambda^a+\theta
k_\lambda^a)J_a} (Z_{k_{1}}^{\delta_{k_1}}\cdots
Z_1^{\delta_1})^\inv Y^{\beta_{j_1}}_{j_1}\cdots
Y^{\beta_{1}}_{1}.
\end{align} We now define an analogous transformation
$\widetilde{Gr}_{t\lambda}$
 that
acts on the holonomy $H_\eta$ by inserting at each intersection
point the vector
\begin{align}
\label{insertel2}
 ((Z_{k_i}^{\delta_{k_i}}
\!\!\cdots\! Z_{1}^{\delta_1})H_\lambda(Z_{k_i}^{\delta_{k_i}}
\!\!\cdots\! Z_{1}^{\delta_1})^\inv)^{\theta t\epsilon_i} =
(Z_{k_i}^{\delta_{k_i}} \!\!\cdots\! Z_{1}^{\delta_1}) e^{-\theta
t\epsilon_i p^a_\lambda J_a}\! (Z_{k_i}^{\delta_{k_i}}
\!\!\cdots\! Z_{1}^{\delta_1})^\inv \in\RR^3\subset\Poi
\end{align}
instead of the Poincar\'e element \eqref{insertel} in the
definition of the Dehn twist.
\begin{align}
\label{grdtdef}\widetilde{Gr}_{t\lambda}:\; H_\eta\mapsto
&Y^{\beta_s}_s\cdots Y^{\beta_{j_n+1}}_{j_n+1}\cdot
(Z_{k_n}^{\delta_{k_n}}\cdots Z_1^{\delta_1}) e^{-\theta
t\epsilon_n(p_\lambda^a+\theta k_\lambda^a)J_a}
(Z_{k_n}^{\delta_{k_n}}\cdots Z_1^{\delta_1})^\inv\cdot
Y^{\beta_{j_n}}_{j_n}\cdots
Y^{\beta_{j_{n-1}+1}}_{j_{n-1}+1}\cdot\nonumber\\
\cdot &(Z_{k_{n-1}}^{\delta_{k_{n-1}}}\!\!\cdots Z_1^{\delta_1})
e^{-\theta t\epsilon_{n-1}(p_\lambda^a+\theta k_\lambda^a)J_a}
(Z_{k_{n-1}}^{\delta_{k_{n-1}}}\!\!\cdots Z_1^{\delta_1})^\inv
Y^{\beta_{j_{n-1}}}_{j_{n-1}}\!\!\cdots
Y^{\beta_{j_{n-2}+1}}_{j_{n-2}+1}()\cdots()\cdot\nonumber\\
\cdot &Y^{\beta_{j_{2}}}_{j_{2}}\!\!\cdots
Y^{\beta_{j_1+1}}_{j_1+1}\cdot(Z_{k_{1}}^{\delta_{k_1}}\cdots
Z_1^{\delta_1}) e^{-\theta t\epsilon_1(p_\lambda^a+\theta
k_\lambda^a)J_a} (Z_{k_{1}}^{\delta_{k_1}}\cdots
Z_1^{\delta_1})^\inv Y^{\beta_{j_1}}_{j_1}\cdots
Y^{\beta_{1}}_{1}.
\end{align}
From the parametrisation \eqref{singlehols} we see directly that
this transformation leaves the Lorentz component of $H_\eta$
invariant and acts on the vector $\bj_\eta$ according to
\begin{align}
\label{grdtjact} \widetilde{Gr}_{t\lambda}:\; &\bj_\eta\mapsto
\bj_\eta+t\sum_{i=1}^n \epsilon_i\Ad(u_{Y_1}^{-\beta_1}\cdots
u_{Y_{j_i}}^{-\beta_{j_i}}) \Ad(u_{Z_{k_i}}^{\delta_{k_i}}\cdots
u_{Z_1}^{\delta_1})\bp_\lambda.
\end{align}

We will now demonstrate that, up to a factor $m_\lambda$, the
transformation \eqref{grdtdef} is the same as the transformation
\eqref{gencurvegraft} of $H_\eta$ under grafting along $\lambda$.
For this, we  express $\lambda$ as a product in the dual
generators $n_{a_i}$, $n_{b_i}$
\begin{align}
\label{doublecurvepar} \lambda=z_t^{\delta_t}\circ\ldots\circ
z_1^{\delta_1}= n_{x_r}^{\alpha_r}\cdot\ldots\cdot
n_{x_1}^{\alpha_1}\qquad x_i\in\{a_1,\ldots,
b_g\},\,\alpha_i\in\{\pm 1\}.
\end{align}
From expression \eqref{dualgens} of $n_{a_i}, n_{b_i}$ in terms of
 $a_i$, $b_i$, it follows that the curves on $S_g$ representing
 $n_{a_i}$, $n_{b_i}$ both start and end in $s_{a_1}$. Hence, by
 representing the curve $\lambda$ as a product of $n_{a_i}$,
$n_{b_i}$, we find that in contrast to the graphical
representation in terms of $a_i$ and $b_i$, there are no
intersection points on straight segments connecting the starting
and endpoints of different factors. All intersection points of
$\lambda$ and $\eta$ occur within the curves representing the
factors $n_{a_i}^{\pm 1}$ and $n_{b_i}^{\pm 1}$ in
\eqref{doublecurvepar}, which reflects the fact that the
generators $n_{a_i}$, $n_{b_i}$ are dual to the generators $a_i,
b_i$. To show that transformation \eqref{grdtdef} agrees with the
transformation \eqref{gencurvegraft} of the holonomy $H_\eta$
under grafting along $\lambda$, it is therefore sufficient to
examine the intersection points of $n_{a_i}$ with $a_i$ and of
$n_{b_i}$ with $b_i$.
Expressing the generators $n_{a_i}, n_{b_i}$  as  products in the
generators $a_i$, $b_i$ via \eqref{dualgens} and applying the
graphical prescription defined above, we find that the
intersection of  $a_i$ and $n_{a_i}$ occurs between $a_i \circ
h_{i-1}\circ\ldots \circ h_1$ and $h_1^\inv\circ\ldots\circ
h_{i-1}^\inv\circ a_i^\inv\circ b_i$ on $n_{a_i}$ and after $a_i$
and has negative intersection number, see Fig.~\ref{inters1}.
Fig.~\ref{inters2} shows that the intersection of $b_i$ and
$n_{b_i}$ occurs before $b_i$ and between $b_i^\inv\circ a_i \circ
h_{i-1}\circ\ldots \circ h_1$ and $h_1^\inv\circ\ldots\circ
h_{i-1}^\inv\circ a_i^\inv \circ b_i\circ a_i$ on $n_{b_i}$, also
with negative intersection number. The intersections of $a_i$ and
$n_{a_i}^\inv$ therefore lie between $b_i^\inv\circ a_i\circ
h_{i-1}\circ\ldots\circ h_1$ and $h_1^\inv\circ\ldots\circ
h_{i-1}^\inv\circ a_i^\inv$ and those of $b_i$ with $n_{b_i}^\inv$
between $a_i^\inv\circ b_i^\inv\circ a_i \circ h_{i-1}\circ\ldots
\circ h_1$ and $h_1^\inv\circ\ldots\circ h_{i-1}^\inv\circ
a_i^\inv\circ b_i$, both with positive intersection number.
 By evaluating the general expression \eqref{grdtjact}
 for the curves
$\eta=a_i,\eta= b_i$, we find
\begin{align}
\bj_\ai\mapsto &\bj_\ai
-\Ad(u_{\ai}^\inv)\!\!\!\!\!\!\!\!\sum_{X_k=\ai,
\alpha_k=1}\!\!\!\!\!\!\!\!\Ad(u_\ai u_{H_{i-1}}\cdots u_{H_1}
v_0^\inv v_{X_{k-1}}^{\alpha_{k-1}}\cdots
v_{X_1}^{\alpha_1}v_0)\bp_\lambda\\
&+\Ad(u_\ai^\inv)\!\!\!\!\!\!\!\! \sum_{X_k=\ai,
\alpha_k=-1}\!\!\!\!\!\!\!\!\Ad(u_\bi^\inv u_\ai u_{H_{i-1}}\cdots
u_{H_1} v_0^\inv v_{X_{k-1}}^{\alpha_{k-1}}\cdots
v_{X_1}^{\alpha_1}v_0)\bp_\lambda\nonumber\\
\bj_\bi\mapsto &\bj_\bi +\!\!\!\!\!\!\!\!\sum_{X_k=\bi,
\alpha_k=1}\!\!\!\!\!\!\!\!\Ad(u_\bi^\inv u_\ai u_{H_{i-1}}\cdots
u_{H_1} v_0^\inv v_{X_{k-1}}^{\alpha_{k-1}}\cdots
v_{X_1}^{\alpha_1}v_0)\bp_\lambda\\
&-\!\!\!\!\!\!\!\! \sum_{X_k=\ai,
\alpha_k=-1}\!\!\!\!\!\!\!\!\Ad(u_\ai^\inv u_\bi^\inv u_\ai
u_{H_{i-1}}\cdots u_{H_1} v_0^\inv
v_{X_{k-1}}^{\alpha_{k-1}}\cdots
v_{X_1}^{\alpha_1}v_0)\bp_\lambda\nonumber,
\end{align}
and with identities \eqref{nprel}, \eqref{helpid1} we recover
\eqref{graftaction}, up to a factor $m_\lambda$. The
transformation of a general curve $\eta\in\pi_1(S_g)$ is then
given by decomposing it into the generators $a_i$, $b_i$, and we
obtain the following theorem
\begin{theorem}
\label{grdttheorem}

Formulated in terms of the holonomies $\ai,\bi$, the grafting map
$Gr_{w\lambda}:(\Poi)^{2g}\rightarrow(\Poi)^{2g}$ defined by
\eqref{graftaction} takes the form
\begin{align}
Gr_{wm_\lambda \lambda}=\widetilde{Gr}_{w\lambda}=D_{\theta
w\lambda},
\end{align}
with $D_{w\lambda}$, $\widetilde{Gr}_{w\lambda}$ given by
 \eqref{dtdef}, \eqref{grdtdef}. In particular, the Poisson bracket between
$m_\lambda$ and $s_\eta$ or, equivalently, $s_\lambda$ and
$m_\eta$ is given by
\begin{align}
\label{dualform} \{m_\lambda, s_\eta\}=\{s_\lambda,
m_\eta\}=-\sum_{i=1}^n \epsilon_i(\lambda,\eta)
(\Ad(u_{Z_{k_i}}^{\alpha_{k_i}}\cdots
u_{Z_1}^{\alpha_1})\hat\bp_\lambda)\cdot(\Ad(u_{Y_{l_i}}^{\beta_{l_i}}\cdots
u_{Y_1}^{\beta_1})\hat\bp_\eta).
\end{align}
\end{theorem}

Hence, we have found a rather close relation between  the action
of infinitesimal Dehn twists and grafting along a closed, simple
curve $\lambda\in\pi_1(S_g)$. The infinitesimal Dehn twist along
$\lambda$ is generated by the observable $m_\lambda s_\lambda$ and
acts on the holonomy of another curve $\eta$ by inserting at each
intersection point $q_i$ the Poincar\'e element
$((Z_{k_i}^{\alpha_{k_i}}\cdots
Z_1^{\alpha_1})H_\lambda(Z_{k_i}^{\alpha_{k_i}}\cdots
Z_1^{\alpha_1}))^{\epsilon_i t}$. Grafting along $\lambda$ is
 generated by the observable $m_\lambda^2$ and
inserts at each intersection point the element
$((Z_{k_i}^{\alpha_{k_i}}\cdots
Z_1^{\alpha_1})H_\lambda(Z_{k_i}^{\alpha_{k_i}}\cdots
Z_1^{\alpha_1}))^{\theta\epsilon_i t}\in\RR^3\subset\Poi$. The
formal parameter $\theta$ satisfying $\theta^2=0$ therefore allows
us to view grafting along $\lambda$ with weight $w$ as an
infinitesimal Dehn twist with parameter $\theta w$.


\section{Example: Grafting and Dehn twists along $\lambda=h_i=[b_i,a_i^\inv]$}
\label{exsect}
\begin{figure}
\vskip .3in \protect\input epsf \protect\epsfxsize=12truecm
\protect\centerline{\epsfbox{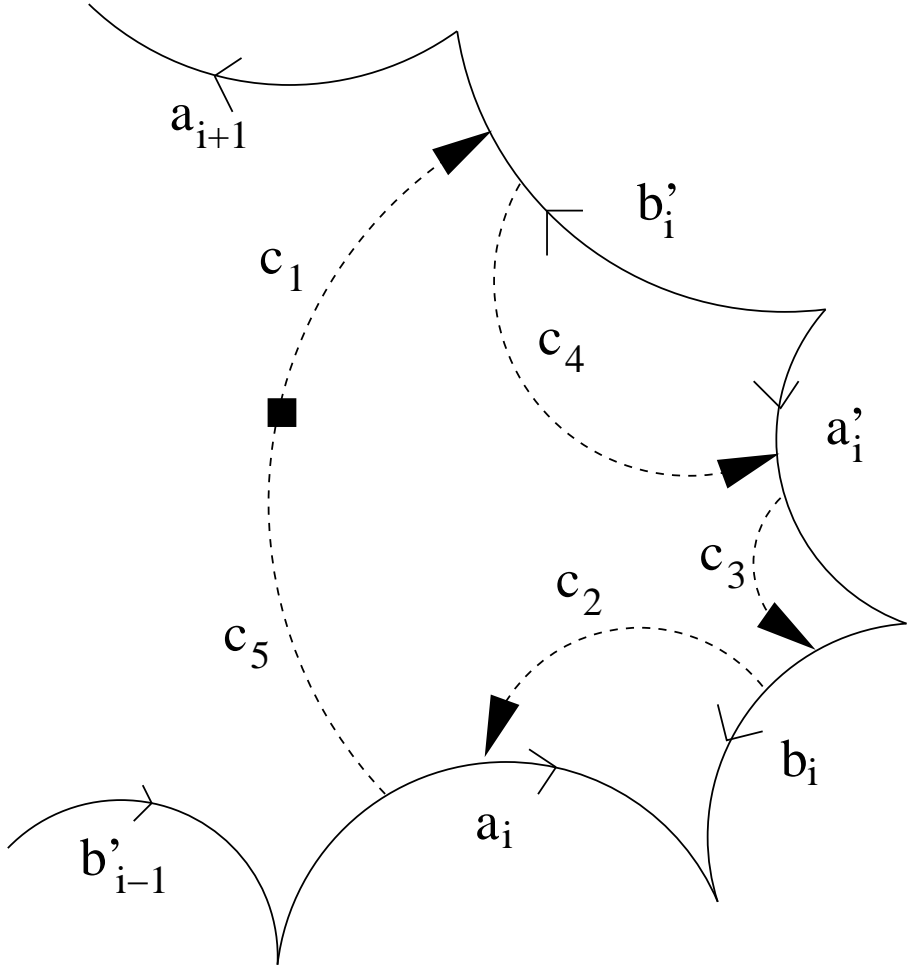}} \caption{The
intersection of the geodesics $\bc_i$ with the polygon
$P_\Gamma^1$} \label{example}
\end{figure}
To illustrate the general results of this paper with a concrete
example, we consider grafting and Dehn twists along the curve
$\lambda=h_i=[b_i,a_i^\inv]\in\pi_1(p_0,S_g)$.

We start by determining the transformation of the holonomies under
grafting along $\lambda$ as described in Sect.~\ref{Csgraft}. From
\eqref{holexp4} it follows that the associated element of the
cocompact Fuchsian group $\Gamma$ is given by
\begin{align}
\label{vel} v=v_0 u_{H_i}v_0^\inv=( v_{H_1}^\inv\cdots
v_{H_{i-1}}^\inv) v_{H_i}^\inv (v_{H_{i-1}}\cdots v_{H_1}).
\end{align}
As we have shown in Sect.~\ref{Csgraft} that conjugation  with
elements of $\Gamma$ does not affect the grafting, we can instead
consider the curve
$$
\tilde\lambda=n_{h_i}^\inv=[n_{a_i}^\inv, n_{b_i}] $$ with
associated group element
$$
\tilde v=v_{H_i}^\inv=[v_\ai^\inv, v_\bi].
$$
We denote by  $\tilde \bc_{p,q}$ the lift of the closed, simple
geodesic $\tilde\lambda$ to a geodesic in $\hyp$ with $\bp\in
P^1_\Gamma$ and with unit normal vector
\begin{align}
\label{exnorm} \tilde{\bn}_{p,q}=-\Ad(v_{H_{i-1}}\cdots
v_{H_1}v_0)\hat \bp_\lambda\qquad e^{-p_\lambda^a J_a}=[u_\bi,
u_\ai^\inv].
\end{align}
From Fig.~\ref{example}, we find that the geodesics in the
associated $\Gamma$-invariant multicurve on $\hyp$ that intersect
the polygon polygon $P_\Gamma^1\subset \hyp$ are given by
\begin{align}
&\bc_1=\tilde \bc_{p,q},\qquad \bc_2=\Ad(v_\bi^\inv)\tilde
\bc_{p,q},\qquad \bc_3=\Ad(v_\ai v_\bi^\inv)\tilde \bc_{p,q}\\
&\bc_4=\Ad(v_\bi v_\ai^\inv v_\bi^\inv)\tilde \bc_{p,q},\qquad
\bc_5=\Ad([v_\ai^\inv, v_\bi])\tilde \bc_{p,q}=\tilde
\bc_{p,q},\nonumber
\end{align}
and all intersection points lie on sides $a_i,a'_i,b_i,b'_i$. The
side $a_i$ of the polygon intersects $\bc_2, \bc_5=\bc_1$ with,
respectively, positive and negative intersection number, while
$b_i$ intersects $\bc_2, \bc_3$, also with, respectively, positive
and negative intersection number. Hence, using formula
\eqref{graftaction} and expression \eqref{exnorm}, we find that
the transformation of the holonomies $\ai,\bi$ along the
generators $a_i,b_i\in\pi_1(S_g)$ is given by
\begin{align} \label{graftex}
\bj_\ai\mapsto &\bj_\ai\!\!-\!\!t\Ad(v_0^\inv
v_{H_1}^\inv\!\!\cdots\!
v_{H_{i-1}}^\inv)\left(\bn_5-\bn_2\right)\!\!=\!\!\bj_\ai\!\!-\!\!t\Ad(v_0^\inv
v_{H_1}^\inv\!\!\cdots\!
v_{H_{i-1}}^\inv)\left(1\!\!-\!\!\Ad(v_\bi^\inv
)\right)\tilde \bn_{p,q}\nonumber\\
=&\bj_\ai+t(1-\Ad(u_\ai^\inv))\hat \bp_\lambda\nonumber\\
\bj_\bi\!\mapsto &\bj_\bi\!\!+\!\!t\Ad(v_0^\inv
v_{H_1}^\inv\!\!\cdots\!\! v_{H_{i-1}}^\inv v_\ai^\inv
v_\bi)\left(\bn_3\!\!-\!\!\bn_2\right)\!\!=\!\!\bj_\bi\!\!-\!\!t\Ad(v_0^\inv
v_{H_1}^\inv\!\!\cdots\!\!
v_{H_{i-1}}^\inv)(1\!\!-\!\!\Ad(v_\ai^\inv v_\hi^\inv))\tilde \bn_{p,q}\nonumber\\
=&\bj_\bi+t(1-\Ad(u_\bi^\inv))\hat \bp_{\lambda},
\end{align}
while all other holonomies transform trivially. The transformation
of the holonomy along a general curve $\eta\in\pi_1(S_g)$ is
obtained by writing the associated vector $\bj_\eta$ as a linear
combination of $\bj_\ai,\bj_\bi$ as in \eqref{etavect}.

\begin{figure}
\vskip .3in \protect\input epsf \protect\epsfxsize=12truecm
\protect\centerline{\epsfbox{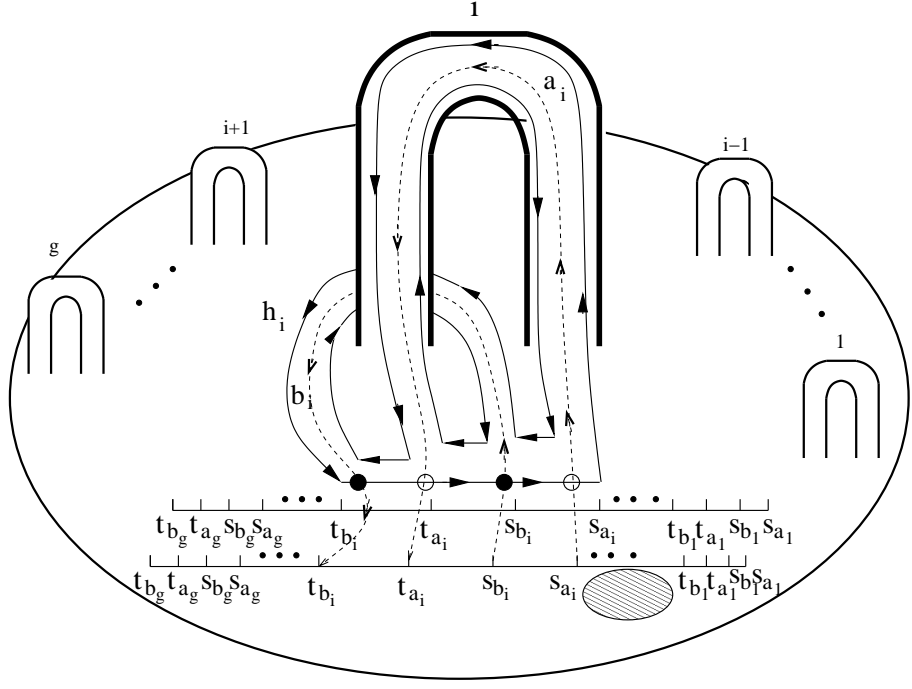}} \caption{The decomposition
 of $h_i$ (full line) and its intersection points with $a_i$, $b_i$ (dashed lines)}
\label{exdt}
\end{figure}
\begin{figure}
\vskip .3in \protect\input epsf \protect\epsfxsize=12truecm
\protect\centerline{\epsfbox{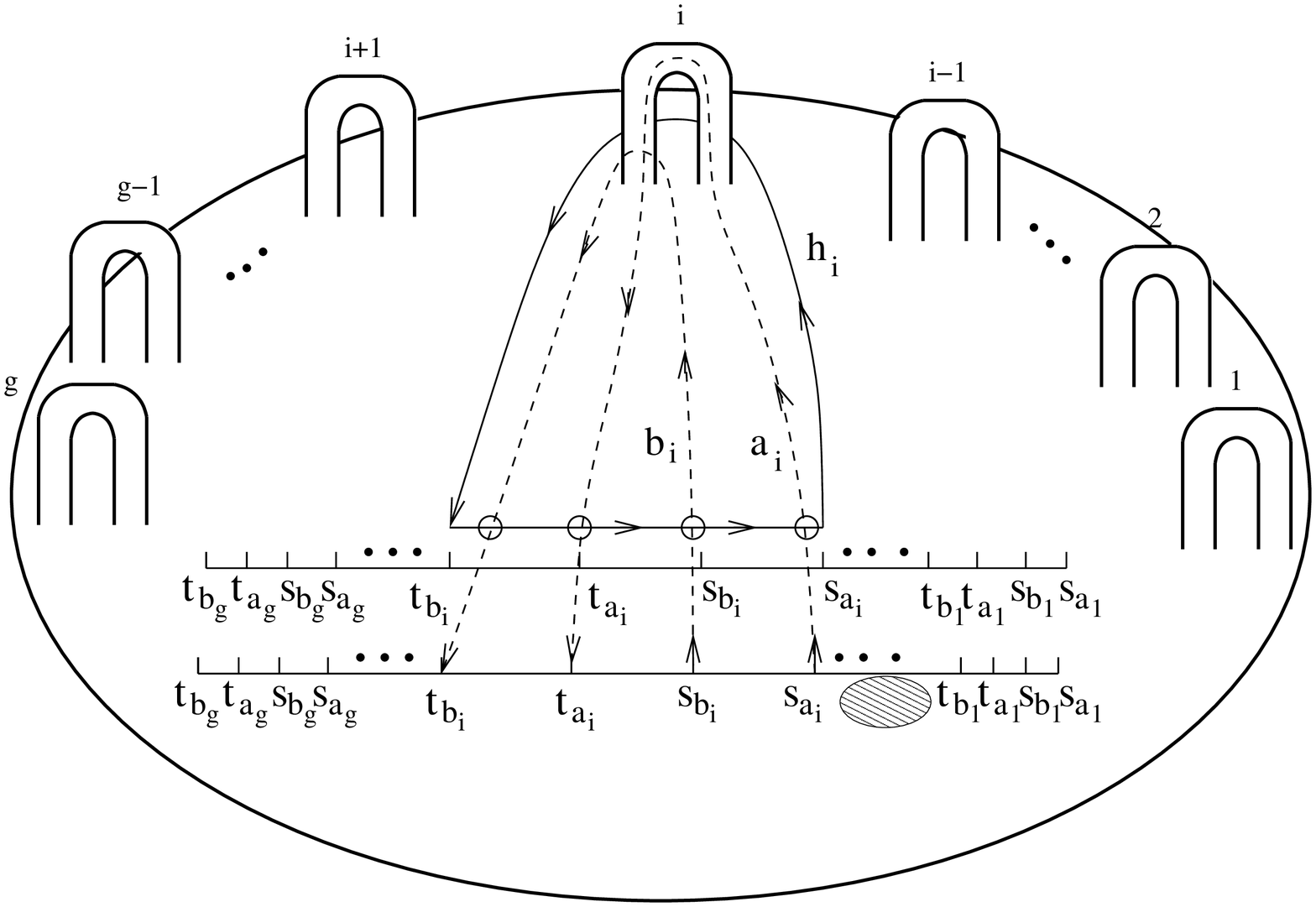}} \caption{The intersection
points of $h_i$ (full line) and its with $a_i$, $b_i$ (dashed
lines), simplified representation without horizontal segments that
do not contain intersection points} \label{exdt2}
\end{figure}
Expression  \eqref{jgrbrackets2} implies that the mass $m_\lambda$
has non-trivial Poisson brackets only with the variables
$\bj_\ai,\bj_\bi$
\begin{align}
&\{ m_\lambda, \bj_\ai\}\!\!= \!\!\Ad(v_0^\inv v_{H_1}^\inv\cdots
v_{H_{i-1}}^\inv)\left( \Ad(v_\bi^\inv)\tilde \bn_{p,q}
\!\!-\!\!\Ad(v_{H_i})\tilde
\bn_{p,q}^c\right)\!\!=\!\!-(1\!\!-\!\!\Ad(u_\ai^\inv))\hat \bp_\lambda\nonumber\\
&\{ m_\lambda, \bj_\bi\}\!\!=\!\!-\Ad(v_0^\inv v_{H_1}^\inv\cdots
v_{H_{i-1}}^\inv v_\ai^\inv v_\bi)\left( \Ad(v_\ai
v_\bi^\inv)\tilde \bn_{p,q}\!\! -\!\!\Ad(v_\bi^\inv)\tilde
\bn_{p,q}\right)\!\!=\!\!-(1\!\!-\!\!\Ad(u_\bi^\inv))\hat
\bp_\lambda\nonumber,
\end{align} Grafting along $\lambda$
therefore acts on the holonomies $\ai,\bi$ according to
\begin{align}
Gr_{tm_\lambda\lambda}:\; &\ai\mapsto (1, t\bp_\lambda)
\ai(1,-t\bp_\lambda)=e^{t \theta p_\lambda^a J_a}\ai e^{-t \theta
p_\lambda^a J_a}=H_\lambda^{-\theta t} \ai H_\lambda^{\theta t}  \\
&\bi\mapsto (1, t\bp_\lambda) \bi(1,-t\bp_\lambda)=e^{t \theta
p_\lambda^a J_a}\bi e^{-t \theta p_\lambda^a
J_a}=H_\lambda^{-\theta t} \bi H_\lambda^{\theta t} .\nonumber
\end{align}
To determine the action of an (infinitesimal) Dehn twist along
$\lambda$, we apply the graphical procedure of Sect.~\ref{grdtrel}
as depicted in Fig.~\ref{exdt}, Fig.~\ref{exdt2}.

We find that both $a_i$ and $b_i$ intersect $\lambda$ twice, once
at their starting points with positive intersection number and
once at their endpoints with negative intersection number. All
intersections take place on the segment linking $t_{b_i}$ with
$s_{a_i}$ on $\lambda$. Hence, the action of an infinitesimal Dehn
twist along $\lambda$ on the holonomies $\ai,\bi$ is given by
\begin{align}
&D_{t\lambda}:\ai\!\mapsto\! e^{t(p_\lambda^a+\theta
k_\lambda^a)J_a}\ai e^{-t(p_\lambda^a+\theta k_\lambda^a)J_a}\!=\!
H_\lambda^{-t} \ai H_\lambda^{t},\;\bi\!\mapsto\!
e^{t(p_\lambda^a+\theta k_\lambda^a)J_a}\bi
e^{-t(p_\lambda^a+\theta k_\lambda^a)J_a}\!=\! H_\lambda^{-t} \bi
H_\lambda^{t},\nonumber
\end{align}
where $H_\lambda=[\bi,\ai^\inv]=e^{-(p^a_\lambda+\theta
k^a_\lambda)J_a}$, and we obtain the relation between grafting and
Dehn twists in Theorem \ref{grdttheorem}:
$Gr_{tm_\lambda\lambda}=D_{\theta t\lambda}$.

\section{Concluding remarks}
\label{conc}

In this paper we related the geometrical construction of evolving
(2+1)-spacetimes via grafting to phase space and Poisson structure
in the Chern-Simons formulation of (2+1)-dimensional gravity. We
demonstrated how grafting along closed, simple geodesics $\lambda$
is implemented in the Chern-Simons formalism and showed how it
gives rise to a transformation on an extended phase space realised
as the Poisson manifold $((\Poi)^{2g},\Theta)$. We derived
explicit expressions for the action of this transformation on the
holonomies of general elements of the fundamental group and proved
that it leaves Poisson structure and constraints invariant.
Furthermore, we showed that this transformation is generated via
the Poisson bracket by a gauge invariant Hamiltonian, the mass
$m_\lambda$, and deduced the symmetry relation $\{m_\lambda,
s_\eta\}=\{s_\lambda, m_\eta\}$ between the Poisson brackets of
mass and spin of general closed curves $\lambda,\eta$. We related
the action of grafting on the extended phase space to the action
of Dehn twists investigated in \cite{we3} and showed that grafting
can essentially be viewed as a Dehn twist with a formal parameter
$\theta$ satisfying $\theta^2=0$.

Together with the results concerning Dehn twists in \cite{we3},
the results of this paper give rise to a rather concrete
understanding of the relation between spacetime geometry and the
description of the phase space in terms of  holonomies. There are
two basic transformations associated to a simple, closed curve
$\lambda$ that alter the geometry of (2+1)-spacetimes, grafting
and infinitesimal Dehn twists. These transformations are generated
via the Poisson bracket by the two basic gauge invariant
observables associated to this curve, its mass $m_\lambda$ and the
product $m_\lambda s_\lambda$  of its mass and spin, and act on
the phase space via Poisson isomorphisms or canonical
transformations.

This sheds some light on the physical interpretation of these
observables. In analogy to the situation in classical mechanics
where momenta generate translations and angular momenta rotations,
the two basic observables associated to a  simple, closed curve in
a
 (2+1)-dimensional spacetime generate
infinitesimal changes in geometry. The grafting operation,
generated by its mass, cuts the surface along the curve and
translates the two sides of this cut against each other. The
infinitesimal Dehn twist, generated by the product of its mass and
spin, cuts the surface along the curve and infinitesimally rotates
the two sides of the cut with respect to each other.

It would be interesting to investigate the relation between
grafting and Poisson structure for other values of the
cosmological constant $\Lambda$ and to see if similar results hold
in these cases. In particular, it would be desirable to understand
if and how the Wick rotation derived in \cite{bb} which relates
the grafting procedure for different values of the cosmological
constant manifests itself on the phase space. Although the
semidirect product structure of the (2+1)-dimensional Poincar\'e
group gives rise to many simplifications, Fock and Rosly's
description of the phase space \cite{FR} can also be applied to
the Chern-Simons formulation of (2+1)-dimensional gravity with
cosmological constant $\Lambda>0$ and $\Lambda<0$. For the case of
the gauge group $SL(2,\CC)$ this has been achieved in \cite{BR,
BNR}. Although the resulting description of the Poisson structure
is technically more involved than the one for the group $\Poi$, it
seems in principle possible to investigate transformations
generated by the physical observables and to relate them to the
corresponding grafting transformations in \cite{bb}.


\subsection*{Acknowledgements}

I thank Laurent Freidel, who showed interest in the transformation
generated by the mass observables and suggested that it might be
related to grafting. Some of my the knowledge on grafting was
acquired in discussions with him. Furthermore, I
 thank  Bernd Schroers for useful
 discussions, answering many of my questions and for proofreading this
 paper.

\end{document}